\documentclass[useAMS,usegraphicx]{mn2e}
\input epsf.tex

\title[Communal Decay]{Secular Effects of Tidal Damping in Compact Planetary Systems} 
\author[Hansen \& Murray] {Bradley M. S. Hansen$^1$\thanks{E-mail:hansen@astro.ucla.edu} \& Norm Murray$^2$\\
$^1$Department of Physics \& Astronomy, University of California Los Angeles, Los Angeles, CA 90095\\
$^2$Canadian Institute for Theoretical Astrophysics, 60 St. George St., University of Toronto, Toronto, ON M5S 3H8, Canada}

\begin{document}

\date{submitted April 2014}

\pagerange{\pageref{firstpage}--\pageref{lastpage}} \pubyear{2014}

\maketitle

\label{firstpage}

\begin{abstract}

We describe the long-term evolution of compact systems of terrestrial planets, using a set of simulations that match the statistical
properties of the observed exoplanet distribution. The evolution is driven by tidal dissipation in the planetary interiors, but the
 systems evolve as a whole due to secular gravitational interactions. We find that, for Earth-like dissipation levels, planetary orbits can be circularised out to
periods $\sim 100$~days, an order of magnitude larger than is possible for single planets. The resulting distribution of eccentricities is
a qualitative match to that inferred from transit timing variations, with a minority of non-zero eccentricities maintained by particular
secular configurations. The coupling of the tidal and secular processes enhance the inward migration of the innermost planets in these systems,
and can drive them to short orbital periods. Resonant interactions of both the mean motion and secular variety are observed, although the
interactions are not strong enough to drive systemic instability in most cases. However, we demonstrate that these systems can easily be
driven unstable if coupled to giant planets on longer period orbits.
\end{abstract}


\section{Introduction}

The evolution of planetary systems takes place over a wide range of timescales. Current theories regarding how planets form
suggest that the first hundred million years of a star's life is a time of great activity. Giant planets form within
a few million years and can then potentially migrate large distances, or interact gravitationally and scatter onto highly
eccentric orbits, leading to possibly tidal capture by the parent star or, in the opposite extreme, ejection from the system
entirely (see Wu \& Murray 2003; Ida \& Lin 2004; Fabrycky \& Tremaine 2007; Naoz et al. 2011; Bromley \& Kenyon 2011; Miguel, Guilera \& Brunini 2011;
Nagasawa \& Ida 2011; Kley \& Nelson 2012; Mordasini et al. 2012 for a representative but incomplete sampling of opinions
on this wide ranging subject). The formation of planets of lower mass is generally held to be more sedate, whether it be by a form of communal
migration or in situ assembly (Raymond, Barnes \& Mandell 2008; Ida \& Lin 2010; Ogihara, Duncan \& Ida 2010; Hansen \& Murray 2012, 2013;
Chiang \& Laughlin 2013; Raymond et al. 2013). Nevertheless, studies in our own solar system indicate that even these nominally
sedate planetary orbits comprise a
chaotic dynamical system (Laskar 1989; Sussman \& Wisdom 1992; Lithwick \& Wu 2011), and that some of the
terrestrial planets may ultimately prove to be dynamically unstable (e.g. Batygin \& Laughlin 2008; Laskar \& Gastineau 2009).

The theme of dynamical stability in planetary systems is therefore of broad interest, and has already attracted significant
attention. The discovery of compact, densely packed planetary systems in recent years has only heightened the interest and
presents several issues regarding the nature of the planets themselves, and whether the currently observed configurations are
sculpted by the original formation or by dynamical processes on longer timescales. Furthermore, many planets have
been discovered close to their parent stars, so that tidal dissipation is expected to rapidly damp eccentricities.

Our goal in this work is to study the long-term evolution of low-mass, compact, multiple planet systems, under the influence
of tidal dissipation, while also accounting for the secular gravitational coupling between the planets.
 This has been a subject of some interest for gas giant planets, which are often found in
strongly hierarchical systems (Wu \& Goldreich 2002; Adams \& Laughlin 2006a; Mardling 2007; Batygin, Bodenheimer \& Laughlin 2009;
Correia et al. 2011; Greenberg \& van Laerhoven 2011; Zhang, Hamilton \& Matsumura 2013). We wish to extend it to the realm of lower mass planetary systems recently
discovered by both radial velocity (Howard et al. 2010; Mayor et al. 2011) and transiting methods (Borucki et al. 2011; Batalha et al. 2013).
 In order to do this, we need to address the issue of large multiplicity and the stronger tidal dissipation anticipated for 
terrestrial class planets. In \S~\ref{Arch} we will describe the initial configurations of the planets under study and our model for tidal dissipation is
given in \S~\ref{Tides}. \S~\ref{Results} then examines the effects of this combination for the long-term structure of these planetary systems.

\section{Secular Architectures}
\label{Arch}

\begin{table*}
\centering
\begin{minipage}{140mm}
\caption{Initial Conditions for a Sample System. Radii are taken
from Seager et al. (2007), assuming the modified Perovskite equation of state. \label{Sample}}
\begin{tabular}{@{}lccccccc@{}}
\hline
 $\#$ &
M & R & a & e  & $\bar{\omega}$ & i & $\Omega$ \\
    &($M_{\oplus}$) & ($R_{\oplus}$) & (AU) &  & $^{\circ}$ & ($^{\circ}$) & ($^{\circ}$) \\
\hline
1 & 2.06 & 1.27 & 0.0616 & 0.124 & 177 & 4.1 & 203 \\
2 & 1.89 & 1.24 & 0.1155 & 0.056 & 146 & 8.4 & 315 \\
3 & 2.21 & 1.30 & 0.1762 & 0.095 & 95 & 0.8 & 1 \\
4 & 3.19 & 1.44 & 0.2858 & 0.131 & 180 & 1.8 & 23 \\
5 & 4.57 & 1.60 & 0.4457 & 0.029 & 346 & 2.8 & 100 \\
6 & 1.07 & 1.05 & 0.6908 & 0.065 & 271 & 1.1 & 130 \\
7 & 1.27 & 1.10 & 0.9459 & 0.138 & 95 & 1.3 & 11  \\
\hline
\end{tabular}
\end{minipage}
\end{table*}

The methods used to uncover low mass planetary systems are subject to systematic errors and selection effects that limit the
completeness with which we can characterise the multiplicity of these planetary systems. This is especially true for modelling
transiting systems, in which only a fraction of the planets may be on transiting orbits at any given time. As a result, many
authors have adopted statistical comparisons with the data to evaluate issues such as compactness, inclination dispersion,
and multiplicity (Lissauer et al. 2012; Fabrycky et al. 2012; Fang \& Margot 2012; Tremaine \& Dong 2012). 

The secular behaviour of a planetary system is determined by the distribution of mass amongst all of its constituent planets,
making an accounting for unobserved components particular important and therefore requiring the adoption of a specific planetary model.
In a previous paper (Hansen \& Murray 2013), we have performed simulations of the in situ assembly of
rocky planetary systems and demonstrated that the resulting period and inclination distributions provide a reasonable comparison
to the statistical properties of the transiting systems observed by Kepler. Thus, we will use
these systems as the initial conditions for our calculations. The end point of each assembly simulation provides a full description
of the masses and semi-major axes of the planets, which will determine the architectures, as well as snapshots of the eccentricities,
inclinations and associated nodal and periastron longitudes, which will determine the initial conditions. Table~\ref{Sample} provides
an example of such a system -- seven  planets within 1~AU, with masses of a few Earth masses each, orbiting a central star of $1 M_{\odot}$.

However, it should be noted that the secular configurations that result from these simulations are not specific to an in situ
assembly scenario, as any origins model that produces similar masses and spacings will have a similar secular behaviour.

\subsection{Secular Modes}
\label{Modes}


The recent interest in secular evolution of planetary systems has seen the development
of a variety of reformulations of the secular planetary  problem (e.g. Veras \& Armitage 2007; Laskar, Boue' \& Correia 2012; Mardling 2013).
These have been largely driven by the observed giant planet systems and have thus been tailored to treatments of higher eccentricities or the
hierarchical nature of the observed giant planet systems.
 The configurations of interest in this paper are closely packed systems
of low mass planets, with mostly low-to-moderate eccentricities and inclinations. As such, their 
secular interactions are mostly well described within the context of the classical Laplace-Lagrange secular perturbation theory (see Murray \& Dermott 1999),
although the compact nature of the systems does require the addition of a term to treat the relativistic precession
of close planetary orbits (e.g. Adams \& Laughlin 2006b). The high multiplicity
of the systems under discussion also leads to a non-negligible frequency of near-commensurabilities amongst
the orbital periods of the modelled systems, which does require the addition of another element to the classical theory. 
To that end, we have incorporated the secular effects of perturbations
due to proximity of neighbouring pairs to the first order 3:2 and 2:1 resonances (Malhotra et al. 1989), as this
can be smoothly incorporated into the traditional secular eigenvalue problem. The same cannot be done for higher
order resonances, but these are also of higher order in the (low) eccentricities and mass ratios considered here and are thus assumed to be negligible.

Thus, our analysis of the secular interactions of a particular planetary system is based on the solution to the eigenvalue problem
\begin{equation}
 A_{ij} x_j = g \, x_i \label{EigenProb}
\end{equation}
where the matrix elements $A_{ij}$ are a combination of the classical secular perturbations and the secular averages of the near resonant
perturbations, given in appendix~\ref{Formulae}, and $g$ is the resulting eigenvalue.

The planetary systems we consider can have as many as 10~planets within 1~AU of the host star, which renders analytic
treatments of the eigenvalue problem prohibitive.
 As such, we determine the eigenvalues numerically. We use standard numerical methods (Press et al. 1992) to reduce each
matrix to Hessenberg form and then solve for the eigenvalues. 
We also determine the eigenfunctions by solving the resulting linear system for each eigenvalue. Convergence was aided by starting with
an initial guess of small amplitude and solving iteratively.

\begin{figure}
\includegraphics[width=84mm]{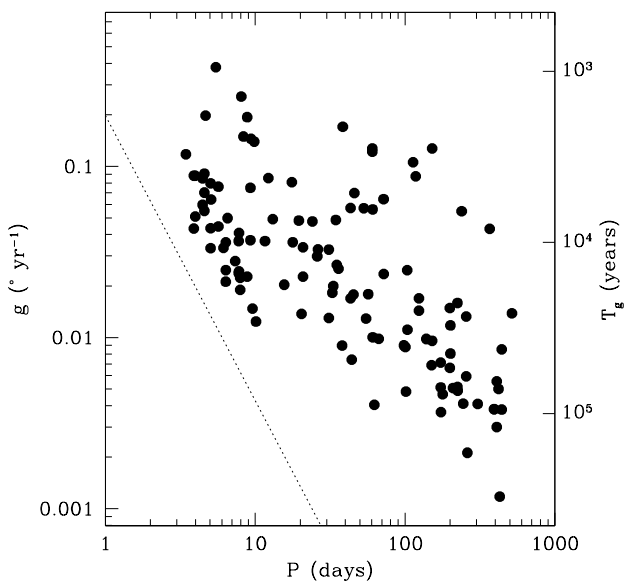}
\caption[gscan.ps]{The points show the eigenfrequencies of the model systems before tidal evolution, plotted against
the orbital period of the dominant planet in the corresponding eigenfunction. The dotted line indicates the value of
the relativistic precession as a function of orbital period.
\label{gscan}}
\end{figure}

This procedure yields numerical solutions of the eigenfrequencies and eigenfunctions of the systems under consideration. Figure~\ref{gscan} shows
the resulting eigenvalues for an ensemble of 20 model planetary systems orbiting a 1$M_{\odot}$ star,
 as a function of the orbital period of the dominant planet in each corresponding eigenfunction. We see that the characteristic
periods of oscillation for the secular modes span the range $10^3$--$10^5$~years, with a general trend that longer periods are
found for modes dominated by more distant planets. 
Interestingly, the secular modes of various
observed giant planet systems studied in Adams \& Laughlin (2006b) also span a similar period range. This commonality of timescales
is dictated by the wider spacing of the more massive systems, to be expected from general considerations of long-term dynamical stability.

The resulting eigenfunctions also demonstrate the degree to which these multiple planet systems are strongly coupled.
If we define two planets as coupled if they both contribute more than 10\% of the amplitude of a given eigenvector,
then we can characterise which planets of a given system are coupled together by the secular modes of oscillation.
Only 44\% of the multiple planet systems here are fully coupled, in the sense that one can construct a chain of
direct coupling between the innermost and outermost planets. In the other cases, the secular modes break down into
two (on rare occasions, three) groups which are only marginally coupled in the sense that none of the inner planets contribute more than 10\%
of the amplitude to any eigenvector dominated by the outer planets, and vice versa. The division usually occurs
between 0.25--0.45~AU, corresponding to orbital periods from 45--110~days.

\section{Evolution under the action of Tides} 
\label{Tides}

The consequence of tidal dissipation in extrasolar planets is to remove energy from the orbit, but not angular momentum.
For a single planet, this means circularisation of the orbit, accompanied by the amount of orbital decay necessary to  
conserve angular momentum. When the planet with tides is in a multiplanet system, the secular modes of the system can
exchange angular momentum and so the tidal evolution of the planet is now coupled to the secular oscillations of the
system as a whole. The long term effect of this coupling is to damp the amplitudes of particular eigenmodes at different
rates (Wu \& Goldreich 2002; Mardling 2007; Batygin, Bodenheimer \& Laughlin 2009; Greenberg \& van Laerhoven 2011; Zhang, Hamilton \& Matsumura 2013). For systems of small multiplicity, this
can often result in a late-time configuration which is dominated by a single mode, characterised by apsidal alignment of
the planetary orbits, and characteristic ratios of orbital eccentricities -- the so-called 'fixed point' evolution.  
The planetary systems under discussion here are richer than the two or three planet systems usually discussed in this context to date.
One question of interest then is how many of the modes of these high multiplicity systems are damped, and how is this
reflected in the planetary orbits?

\subsection{Tidal Dissipation Calibration}

To answer such questions, we also require
  a physical model for the tidal dissipation if we want to examine the evolution of a model system.
 We have previously calibrated the strength of tidal interactions in giant
extrasolar planets (Hansen 2010) but cannot realistically apply the same normalisation to the possibly very different
interior structure of the rocky planets assumed to be produced in these simulations. Nevertheless, we can
adopt the same formalism as long as we use a different astrophysical normalisation. The traditional normalisation
of Earth-like planets is using the secular acceleration of the Moon, and expressed in terms
of the ``modified tidal quality factor'' $Q'$. This reflects the number of forcing periods 
required to dissipate the tidal potential energy,
 and is estimated for the Earth to be 
$Q'_{\oplus} \sim 10$ (Goldreich
\& Soter 1966). If we express the tidal $Q'$ factor in terms of the bulk dissipation $\sigma_{\oplus}$ using
the same formalism as in Hansen (2010), we derive that
$$
\sigma_{\oplus} \sim 3.2 \times 10^{10} \left( \frac{Q'_{\oplus}}{10} \right)^{-1} \sigma_J
$$
where $\sigma_J$ is the bulk dissipation constant we previously derived for giant planets. Thus, the dissipation per unit mass in
rocky bodies is considerably larger than in gaseous planets. Renormalising the tidal evolution equations
of Eggleton et al. (1998) using this result, and truncating the eccentricity beyond quadratic order (since
we are concerned with small eccentricities in this paper), we find
\begin{eqnarray}
\frac{d \ln a}{dt} & = & \frac{e^2}{3.14 \times 10^{9} yrs} \left( \frac{a}{0.1 AU} \right)^{-8} \left( \frac{R_p}{R_{\oplus}} \right)^{10} \frac{M_{\oplus}}{M_p} \label{Tideqa} \\
\frac{d \ln e}{dt} & = & \frac{ (1 + 56.5 e^2)}{1.7 \times 10^{10} yrs}  \left( \frac{a}{0.1 AU} \right)^{-8}  \left( \frac{R_p}{R_{\oplus}} \right)^{10} \frac{M_{\oplus}}{M_p} \label{Tideq}
\end{eqnarray}
This latter expression serves as the effective eccentricity damping rate $1/\tau$ for each planet in our discussion of \S~\ref{Numerics}, below.
Furthermore, we shall assume rocky planets, and therefore adopt a mass-radius relation for Perovskite, from Seager et al. (2007).
The planetary eccentricities oscillate due to the secular interactions, so the tidal evolution is calculated by averaging
 the quantity $e^2$ over the secular eccentricity oscillation for each planet.

Given the unknown nature of the dissipation in extrasolar terrestrial planets, the Earths estimated
$Q'_{\oplus}=10$ is a convenient, but very uncertain, benchmark. The physical origin of Earths tides is
held to be some combination of dissipation in the boundary layers of marginal seas (e.g. Jeffreys 1921)
and dissipation due to pelagic turbulence (generated by topographic features) from ocean floors (e.g. Bell 1975; Egbert \& Ray 2000).
Dry planets are likely to have $Q'$ an order of magnitude larger, as has been estimated 
 for Mars, based on the orbit of Phobos 
 (Lainey et al. 2007). On the other hand, the presence of substantial Hydrogen envelopes on many observed extrasolar
rocky planets may enable the propagation of gravity waves and so enhance the dissipation that results from the conversion of barotropic flows into baroclinic waves
 by topographic features (e.g. Bell 1975; Balmforth, Ierley \& Young 2002), especially if hotter planets exhibit
greater mantle plasticity (see discussion in Henning, O'Connell \& Sasselov 2009).

 As such, we will use the above dissipation
as our benchmark and consider variations by an order of magnitude in either direction in \S~\ref{Q}. 
 

\subsection{Numerical Implementation}
\label{Numerics}

To incorporate tidal effects into the secular evolution, we adopt an approach similar to that of Greenberg \& van Laerhoven (2011), with the
generalisation that we include an arbitrary number of planets and allow for tidal dissipation in all planets
simultaneously. 
The inclusion of the eccentricity damping for each planet means that
 each diagonal term $A_{jj}$ in the secular matrix acquires an additional
imaginary contribution $i/\tau_j$, where $\tau_j$ is the eccentricity damping timescale for planet j, calculated from equations~(\ref{Tideq}).
 The inclusion of such a term in the secular equations introduces an imaginary
contribution to the eigenvalues (\ref{EigenProb}).
 We therefore solve the complex N$\times$N matrix as a 
real 2N$\times$2N matrix, which yields the oscillation frequencies in the real part of the solution and the 
mode damping rates in the imaginary part. We evaluate the resulting eigenvectors using the real part of the
eigenvalues only, because the resulting matrices are close to singular (by virtue of the small values of the
imaginary part). In effect, the real part of our solution yields the instantaneous secular eigenmodes at each timestep and
the imaginary parts drive a slow evolution.

The secular solution is then stepped forward by a timestep $\Delta t$. The default value is $\Delta t =1$~Myr,
subject to the condition that this
 be larger than the
largest period of oscillation of the eigenmode spectrum (which is always the case here, see Figure~\ref{gscan})
and substantially smaller than the characteristic tidal dissipation time (also ubiquitously satisfied).
The evolution of the secular solution is then given by 
\begin{equation}
h_i(t+\Delta t) = h_i (t) + \sum_{j=1}^N E_{ij} e^{-\gamma_j \Delta t} \sin \left( g_j \Delta t + \beta_j \right)
\label{hevol}
\end{equation}
and similarly for the $k_i$. In the above, $E_{ij}$ represent the amplitudes of the individual eigenmode contributions
to $h_i$, $\gamma_j$ is the damping rate of mode j calculated from the eigenvalue solution, and $\beta_j$ are the
phase offsets for mode j. Damping of the eccentricity must also be accompanied by evolution in the semi-major axis
(by angular momentum conservation), and the resulting change 
 of the semi-major axis is calculated using the tidal evolution
equation~(\ref{Tideqa}). 
The evaluation of the expressions in equations~(\ref{Tideqa}) and (\ref{Tideq}) require an integration over the
secular oscillations in eccentricity, which we take as 
$<e_i^2>=<h_i^2+k_i^2>$, and which is evaluated using the equation~(\ref{hevol}).

As noted by Greenberg \& van Laerhoven (2011), the evolution in semi-major axis 
 results in a small shift in the properties of the secular solution on top of the eccentricity damping. Thus, after
 each timestep, the new orbital configuration and eccentricities are
 used as initial conditions to solve the eigenvalue problem again for the (slightly) changed eigenvalue and eigenmodes, thereby
updating the $E_{ij}$ and $\beta_j$ in response to the change in semi-major axis, and for calculating a new set of damping
rates $g_j$.

\begin{figure}
\includegraphics[width=84mm]{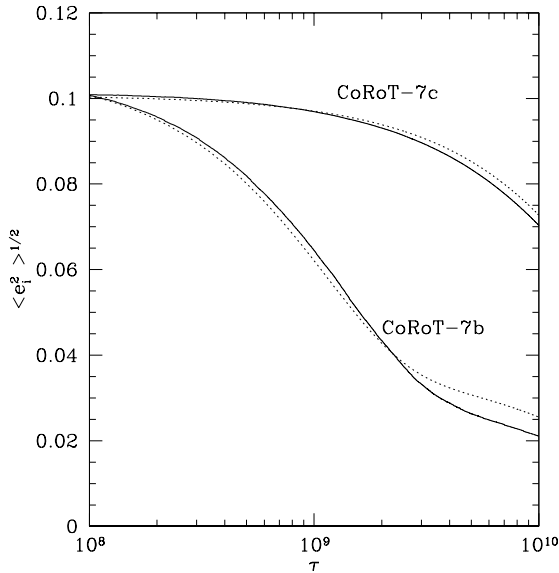}
\caption[GvL.ps]{The solid lines are the tidal evolution for our simple model of the CoRoT-7b system as modelled
by Greenberg \& van Laerhoven (2011). The dotted lines are the solutions derived from numerically evolving their
equations (22). The actual value of $\tau$ is arbitrary, and the two solutions have been scaled so that they match
at the upper left. Deviations at late times are likely the result of the fact that our model accounts
for the evolution in semi-major axis beyond accounting for the Taylor expansion at early times.
\label{GvL}}
\end{figure}

 As a demonstration of
the method, Figure~\ref{GvL}
compares our evolution with the analytic solution (equations 22) of Greenberg \& van Laerhoven (2011) for the two planet system
CoRoT-7b. For this case, we adopt a simpler tidal dissipation (no scaling with semi-major axis or mass)  than that used above, because we wish to approximate
the analytic solution more accurately. Similarly, we neglect the near-resonant terms (which are negligible in this case anyway).
 In Figure~\ref{GvL}, the solid line shows our numerical evaluation, while the dotted
line shows the analytic solution. The character of the evolution is well matched, although there are small deviations at late times
due to the fact that the analytic solution does not account for changes in the secular architecture beyond the first
derivative in the initial conditions.

\subsection{Resonant Crossings}

Tidal evolution can result in the decrease of the semi-major axis for one or more of the planets, with a consequent evolution
in period ratios between neighbouring planets. This can result in the tidally-induced crossing of resonances, which will lead
to divergences in the near-resonant contributions to the secular perturbations given in appendix~\ref{Formulae}.

Resonant interaction can be an exceedingly complex phenomenon, but our particular application is simplified in part by
the fact that the tidal evolution leads to the divergence of orbits, which means that pairs of planets are not captured
into resonance. Furthermore, in the limit of low eccentricities, the tidal evolution through such resonances is rapid
(Lithwick \& Wu 2012; Batygin \& Morbidelli 2013, hereafter LW12 and BM13). Thus, we shall adopt the following approximation, applicable for the
diverging orbits we encounter here.

When the period ratio of a pair of orbits comes within $\epsilon$ of a 2:1 or 3:2 commensurability from below, we assume that the
semi-major axes instantaneously shift so as to move the pair to a period ratio that is $\epsilon$ above the commensurability,
while assuming that total angular momentum is conserved and individual eccentricities are conserved. This transformation results
in a relationship between $a_1$ (the semi-major axis of the inner planet before encounter) and $a'_1$ (the value after encounter),
that is
\begin{equation}
\frac{a'_1}{a_1} = \left(\frac{\left[m_1 \sqrt{1-e_1^2} + m_2 \sqrt{(2-\epsilon) (1-e_2^2)} \right]}
{\left[m_1 \sqrt{1-e_1^2} + m_2 \sqrt{(2+\epsilon) (1-e_2^2)} \right]}    \right)^2,
\end{equation}
for the case of the 2:1 commensurability, and correspondingly for others. The proximity to resonance $\epsilon$ where this occurs
is chosen to be  the larger of the two libration amplitudes calculated for the circular restricted problem with
a test particle and either an inner or outer massive perturber (see Murray \& Dermott 1999 for a detailed discussion of
the latter case). These are also given in the appendix~\ref{Formulae} for the cases of the 2:1 and 3:2 resonances.

We have compared our results to direct numerical integrations of sample systems, using the Mercury integrator (Chambers 1999).
The tidal force is implemented using the radial force expression of Hut (1981). The slow rate of tidal evolution makes
numerical integration with realistic dissipation impractical
so we integrate with a tidal dissipation rate that is 
artificially high by a factor of 150, i.e. an effective $Q'=0.07$ and compare to our secular model with a similarly
high dissipation rate.

\begin{figure}
\includegraphics[width=84mm]{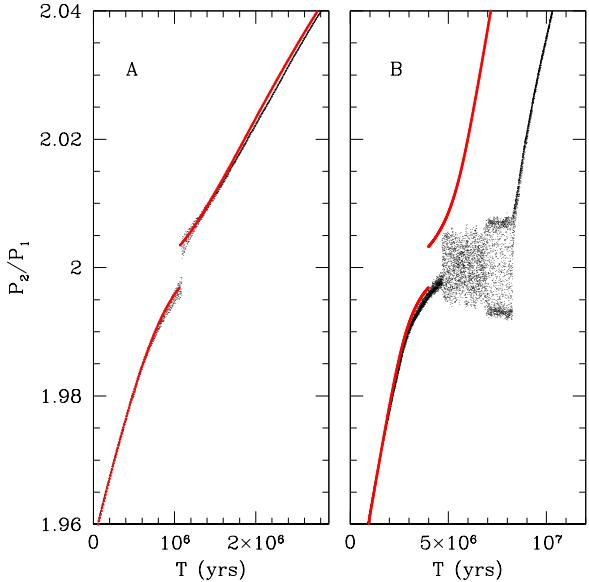}
\caption[Merc9comp.ps]{The two panels show the passage of planet pairs from two different simulations through the 2:1 resonance. In each
case the black points represent a direct numerical integration and the red points represent our secular evolution model. The left panel shows
that our model accurately captures the evolution when passage through the resonance is rapid, but the right panel shows that the passage through
the resonance can follow a more extended path. However, the overall slope of the evolution is approximately recovered once the pair is no
longer in resonance.
\label{Merc9comp}}
\end{figure}

\begin{figure}
\includegraphics[width=84mm]{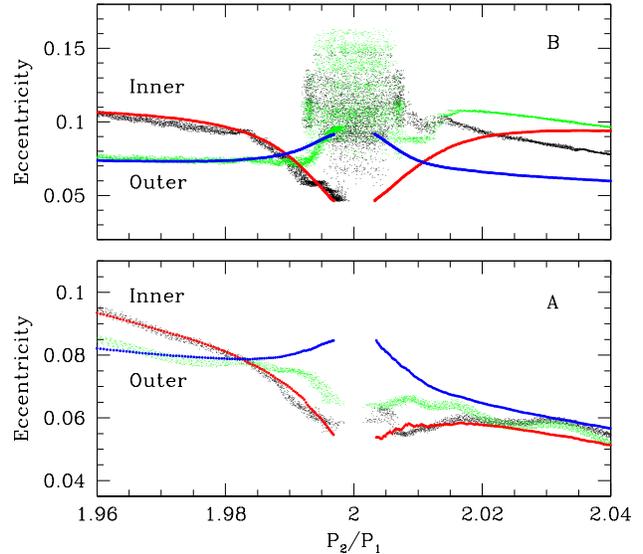}
\caption[MercEcc.ps]{In each panel, the black and green points represents the eccentricities of the inner and outer planet respectively, as determined
by direct numerical integration. The plotted results represent a running boxcar average over $10^5$ years, to remove the fluctuations from higher frequency
terms.
The red and blue curves indicate the corresponding eccentricities as determined by our secular evolution model. We
see that the general character of the evolution is well captured as the period ratio evolves. The fact that the deviations are larger for the outer
planet in each case suggests that additional unmodelled perturbations from other planets in the system are not entirely negligible.
\label{MercEcc}}
\end{figure}

Figure~\ref{Merc9comp} shows the evolution of two planet pairs (the inner pairs of  6 and 7 planet systems respectively) that pass through the 2:1 resonance. The black points
show the full numerical integration and the red curves represent the model evolution. The evolution prior to resonance
crossing is well captured by our model. In the left hand panel, we see that the system crosses the resonance very
rapidly, displaying the form of evolution termed `resonant repulsion' by LW12 and the 
 model reproduces the rapid transition. However, we note that  not all systems follow this evolution, and this is shown
in the right hand panel. Although diverging orbits are not subject to long term resonant capture, some can get delayed in resonance
for a finite period of time. This occurs when the splitting between the different resonant terms in a given commensurability is
larger than their individual widths. This violates the assumption of zero free eccentricity adopted by LW12/BM13. We have opted to not
complicate our simple model further to 
 capture this dynamics because the capture is transient and does not significantly influence the long term evolution. This can be 
seen in Figure~\ref{MercEcc}, which shows the corresponding evolution of the eccentricities in these two cases. Once again, we see that
the long term character of the evolution is well represented, although there are deviations during the resonance passage.
These results give us confidence that our evolutionary model captures the long-term secular behaviour of the systems, although
they suggest timescales could be uncertain at the level $\sim 10^8$ years.

\section{Results}
\label{Results}

We have used the results (after 10 Myr of evolution) of 50 model realisations of the 20$M_{\oplus}$ rocky planet systems from
Hansen \& Murray (2013) to define the initial state of our systems. We assume all the planets are of terrestrial class, in the
sense that they obey the tidal dissipation outlined in \S~\ref{Tides}, and evolve them for 10~Gyr according to our model for
tidal+secular evolution.

\begin{figure}
\includegraphics[width=84mm]{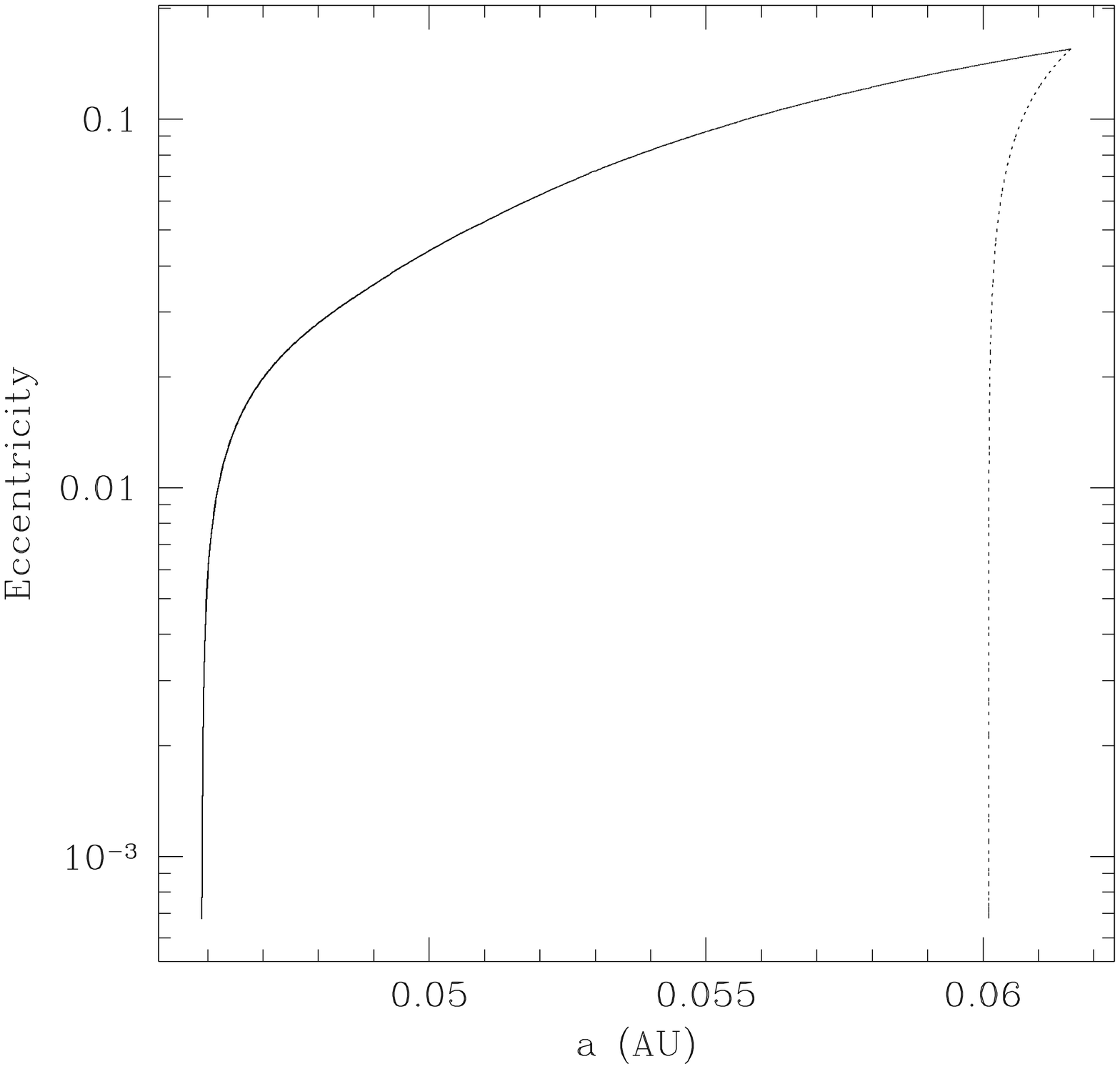}
\caption[Aevol.ps]{The points show the instantaneous semi-major axis and eccentricity of the innermost planet, planet~1.
We see that the eccentricity decreases as the planet moves inwards, but more slowly than expected by the constant angular
momentum curve expected for single planet evolution, which is shown by the dotted line. The pumping of the eccentricity by
secular interactions enables an inward shift approximately four times as large as that experienced by a single planet.
\label{Aevol}}
\end{figure}

The generic consequence of our model is that the inner planet of the system experiences, by far, the largest inward migration.
Furthermore, the continuous excitation of eccentricity due to secular perturbations sustains the tidal dissipation for longer
than it would for a single planet, and consequently drives them in further than would otherwise be the case.
Figure~\ref{Aevol} shows the evolution for the system in Table~\ref{Sample}, which demonstrates this. Such behaviour
was anticipated by previous studies, although we do not find sufficient migration to threaten the survival of any of
our model planets (Mardling \& Lin 2004), because the masses involved are too low to excite eccentricities to the required level.

\begin{figure}
\includegraphics[width=84mm]{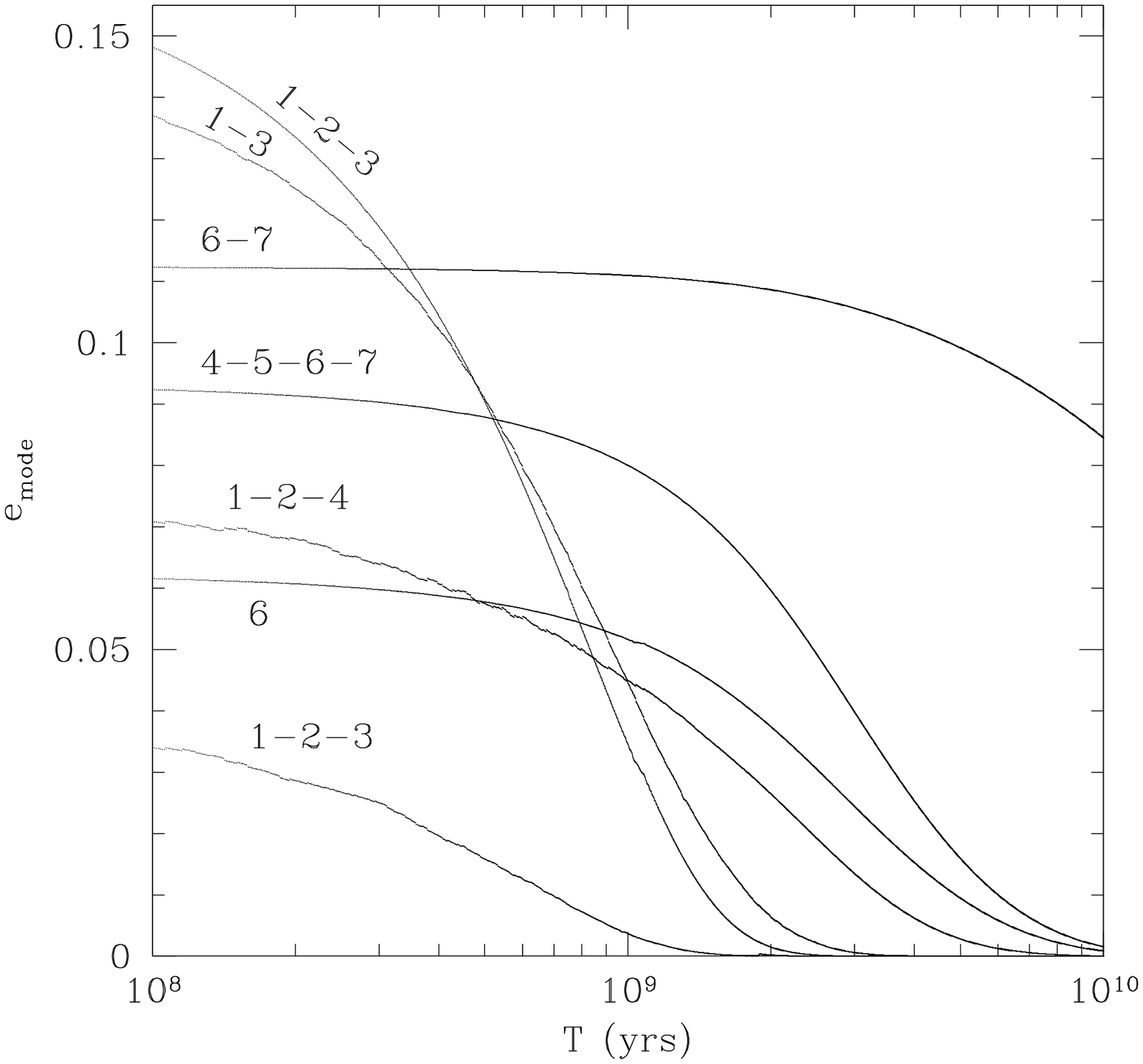}
\caption[Amps.ps]{The numerical labels indicate which planets (counting from inside out) contribute 10\% or more
of the amplitude to a given eigenvector. Thus, we see that the models dominated by the innermost planets (1,2,3) damp
the fastest, while the modes of the outer planets (4,5,6,7) damp on longer timescales. After 10~Gyr, only one mode of
oscillation retains any significant amplitude.
\label{Amps}}
\end{figure}

\begin{figure}
\includegraphics[width=84mm]{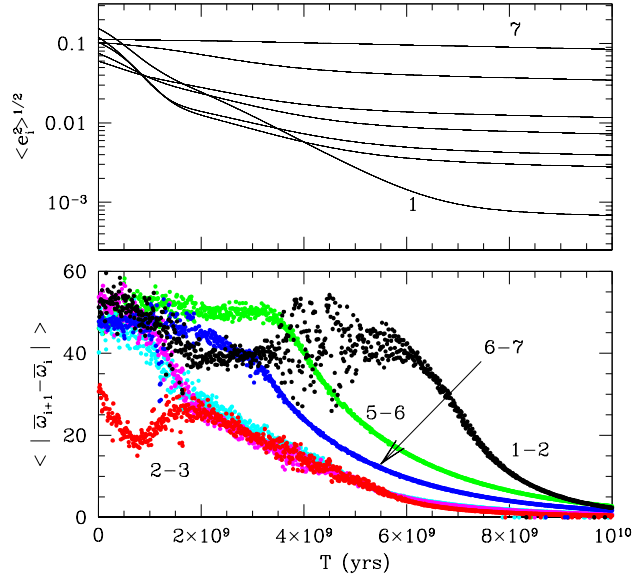}
\caption[te.ps]{The curves in the upper panel indicate the eccentricities of
the individual planets. The curves can be identified with individual planets by
noting that, at late times, the eccentricities increase monotonically with distance
from the star, so only the inner (1) and outer (7) planets are explicitly identified.
 The lower panel indicates the evolution
of the orbital alignment between neighbouring planets. 
The  coloured points correspond to the average offsets in longitude of periastron
for each planet, relative to the planet exterior to it.
 The average is taken in steps of 0.01~Gyr, which
leads to some coarseness for those planets affected by multiple modes.
Colours are black (1-2), red (2-3), magenta (3-4), cyan (4-5), green (5-6) and blue (6-7).
Somewhat surprisingly, it is the innermost pair that takes the longest to come into
alignment. This is a consequence of the fact that the innermost planet, although the most
affected by tides, is also strongly coupled to three other planets. In particular, the
coupling to planet~4 takes several 
Gyr to damp away.
\label{te}}
\end{figure}

Another way to view this evolution is to consider the evolution of the amplitudes
of the various eigenmodes of the system. By virtue of the interaction between the
planets in the system, the tidal damping systematically reduces the amplitude of
the eigenmodes, as determined by the relative contribution each receives from those
planets being affected most strongly by the tides (Wu \& Goldreich 2003; Greenberg \& van Laerhoven 2011).
Figure~\ref{Amps} shows the evolution of the amplitudes of the seven eccentricity
eigenmodes for the system shown in Table~\ref{Sample} and  in Figure~\ref{Aevol}. The secular architecture of this
system is such that it is best characterised as containing subgroups of eigenmodes, which damp on similar
timescales. The fastest damping mode is dominated by planet~2, with significant contributions from 1 and 3, and decays with 1~Gyr.
The two next fastest decay on a timescale $\sim 2$~Gyr, while all but one are substantially damped on timescales $\sim 10$Gyr.
 The equivalent evolution in the planetary rms eccentricity is shown in Figure~\ref{te}, with the closest planets having
the lowest final eccentricities. Note also that the various curves cross during the first several~Gyr when individual planets
are affected by multiple modes, but settle into scaled versions of each other at late times. This is what is expected for a
system which has been tidally damped to the point where only a single mode of oscillation remains excited.
 The lower panel also demonstrates 
this evolution by showing the apsidal alignment of the neighbouring planet pairs. After $10^{10}$ years,
all the apsidal angles are aligned.

\begin{figure}
\includegraphics[width=84mm]{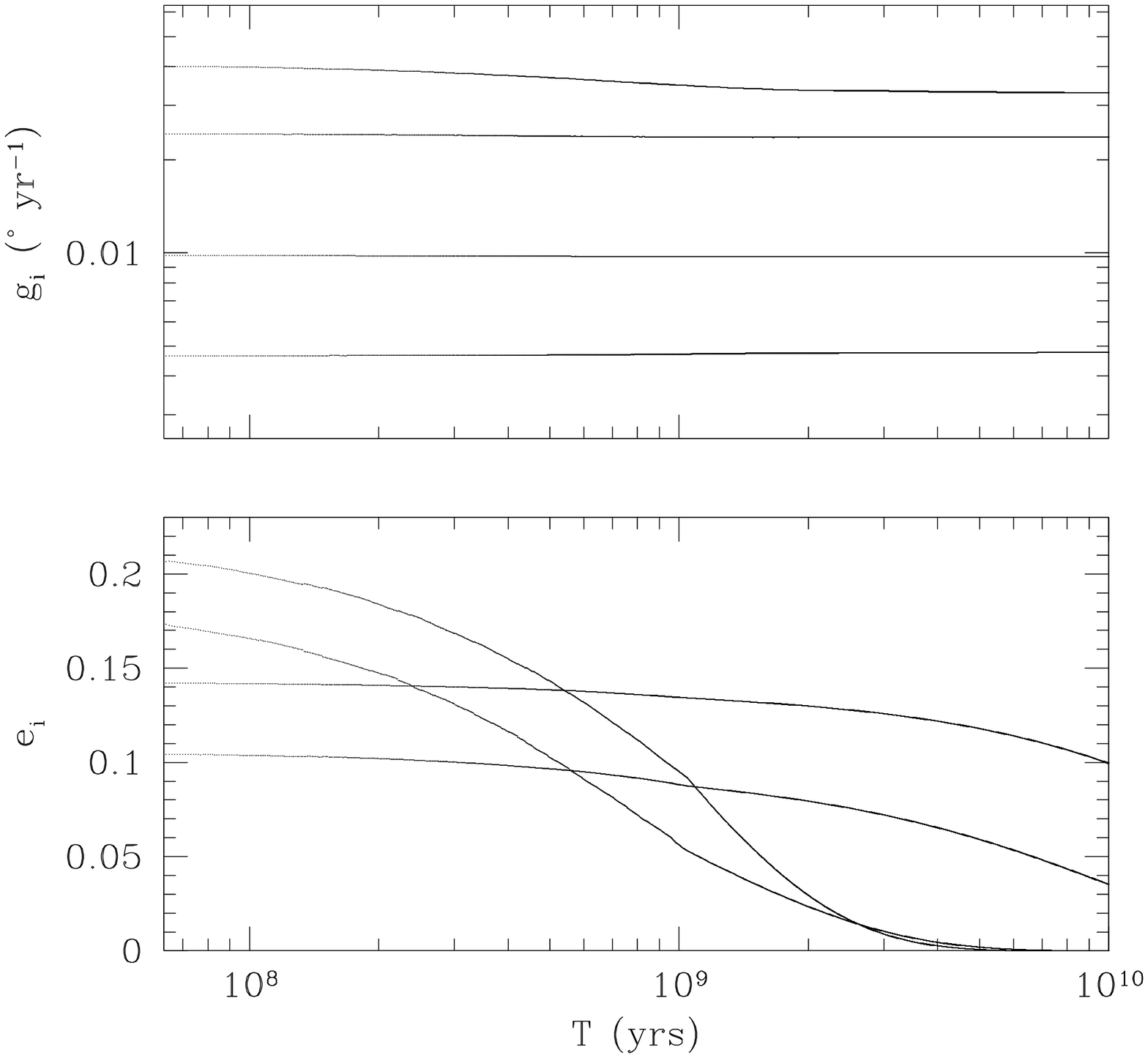}
\caption[Snap1.ps]{The upper panel shows the evolution of the eigenvalues for a four
planet system evolving under the action of planetary tides. The lower panel shows the
corresponding evolution of the amplitudes for the corresponding eigenmodes. We see
that two of the eigenmodes are strongly damped, while two retain non-negligible amplitude
on timescales $\sim 10^{10}$ years.
\label{Snap1}}
\end{figure}

When we review the evolution of the full ensemble of 50 systems, we see a variety of behaviours.
Figure~\ref{Snap1} shows the evolution of the eigenfrequencies and eigenvector amplitudes  for an
evolution of the most common type. It shows the generic behaviour in which the
eigenfrequencies, most notably the highest values, drift to slightly lower values due to the inward shift of the inner planet
semi-major axis. The evolution of the eigenmodes once again demonstrates the tendency for
our multiple planet systems to partially decouple -- the two eigenmodes with significant
contributions from the innermost planet  damp down to almost zero on Gyr timescales, while
the two eigenmodes dominated by the outer planets in this four-planet system evolve much
more slowly. 

\begin{figure}
\includegraphics[width=84mm]{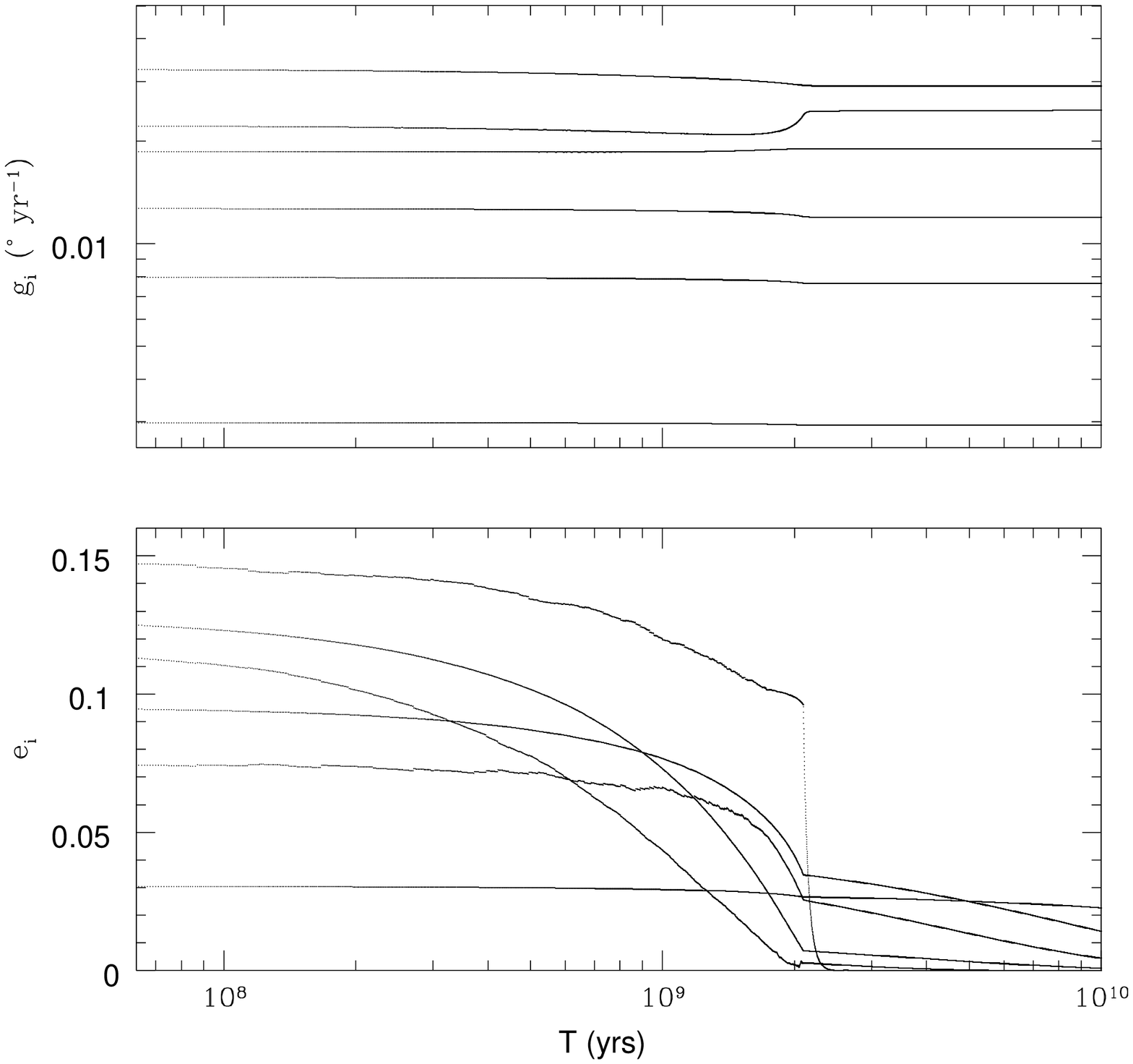}
\caption[Snap2.ps]{The upper panel shows the evolution of the eigenvalues in a six 
planet system where the inner planet becomes largely decoupled from the rest of the system
at about 2~Gyr.
The decoupling of the inner planet leads to the dramatic drop in amplitude for the corresponding mode,
shown in the lower panel, because the eccentricity is no longer excited by secular interactions.
There is also a corresponding change in slope for the evolution of the remaining
modes, which are no longer coupled to that planet.
\label{Snap2}}
\end{figure}

Figure~\ref{Snap2} shows the evolution of a system that is a
 minority, but not uncommon, occurrence ($\sim$ 30\% of the time).
This shows two eigenvalues starting to approach each other but then abruptly flattening out,
 which reflects a situation in which the inner planet is initially coupled to multiple modes
of oscillation but effectively decouples entirely from all but one of the modes as it shifts
inwards due to tidal evolution. When this occurs, the secular pumping of the eccentricity becomes
negligible, the orbit circularizes and the system is now driven by the weaker tidal evolution 
of the second closest planet (hence the flattening of
the curves). The reason for the increase in the second largest eigenfrequency is that this is
the mode dominated by the inner planet, and which is increasingly dominated by the relativistic
precession as it moves inwards.

The effects of the decoupling is also shown to dramatic effect in the mode amplitudes. The loss
of secular eccentricity pumping leads
to a rapid decrease in the amplitude of the mode that has become dominated by the inner planet.
The decoupling also leads to a change in slope of the evolution of the  remaining modes of non-zero amplitude. These
modes do not completely stop evolving because the action of tides on the innermost coupled planet
(which is now the second closest to the star) continues to operate. 

\begin{figure}
\includegraphics[width=84mm]{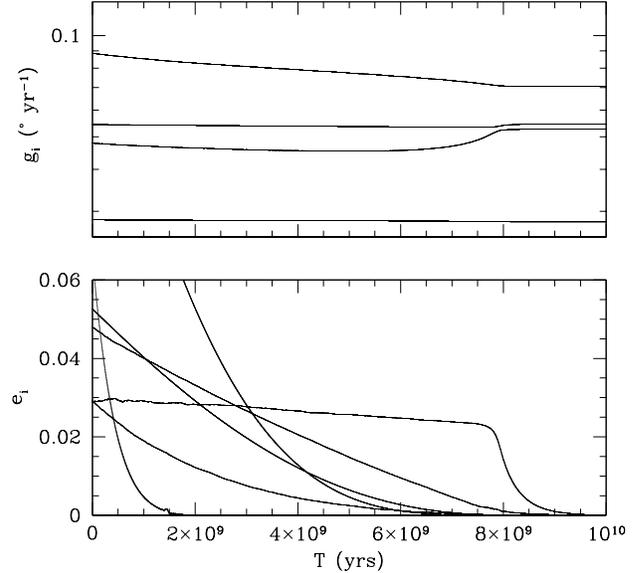}
\caption[Snap3.ps]{The upper panel shows the evolution of four of the eigenvalues for a six
planet system in which the planets are strongly coupled across the entire system (the largest
and smallest frequencies are largely unaffected and off the top and bottom of this plot
respectively). We see the
effect of a secular resonance at $\sim 8$~Gyr, which causes a rearrangement of the
eigenfunctions and changes the rate of evolution of the eigenmode amplitudes in the bottom
panel. In particular, the observed 'knee' is the result of the decoupling of the inner planet
from an outer planet, which causes that mode to decay more rapidly. 
\label{Snap3}}
\end{figure}

Figure~\ref{Snap3} shows that
 even more dramatic behaviour occurs occasionally in which this evolution actually
leads to a secular resonance crossing. This system shows the same
characteristic behaviour as in Figure~\ref{Snap2}, in which two eigenfrequencies approach
each other. However, in this instance, the approach is not resolved by the decoupling of
one planet from the others, and the system undergoes an avoided crossing. This leads to
a rearrangement of the eigenfunctions and a consequent change in the slope of the evolution.
 Similar phenomena have been discussed before
in the context of our solar system, where evolution of the protosolar nebula and the  giant planet system can sweep eigenfrequencies
through the inner solar system (Ward 1981). In our case, the phenomenon appears
to be associated with systems in which a given mode strongly couples two planets that are widely
seperated. In the example shown here, the eigenmode in question couples the inner planet
(at 0.0594 AU initially) with the fifth planet (at 0.4595 AU). The resonance crossing effectively
destroys this coupling, so that the mode becomes dominated by the inner planet, which explains
the more rapid evolution at late times. It is also worth noting that, in this case, the entire system
of planets represents a strongly coupled system, since all six eigenmodes lose substantial amplitude
during the course of the evolution. Indeed, at late times, this entire system exhibits the anticipated
fixed point behaviour.

\begin{figure}
\includegraphics[width=84mm]{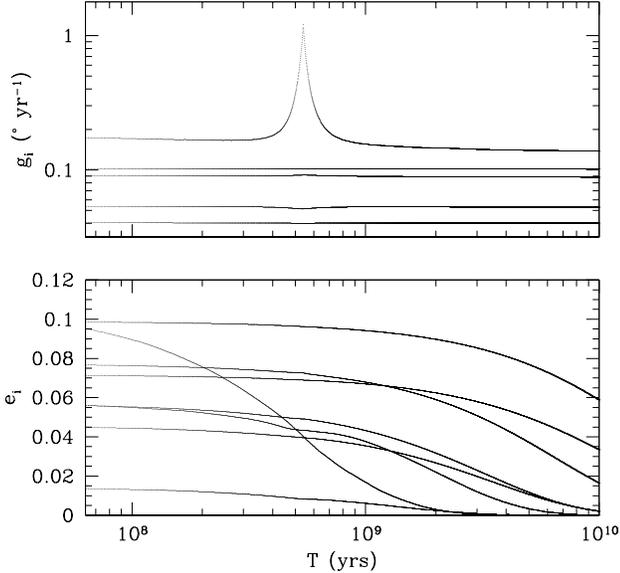}
\caption[Snap4.ps]{The upper panel shows the evolution of the five highest eigenfrequencies of a eight
planet system in which the inner pair crosses the 2:1 resonance at $\sim 525$Myr. When the
system is near resonance, the secular effects of the averaged resonant interaction substantially
alter the eigenfrequencies of those modes with strong contributions from the resonantly interacting
pair. In the lower panel, we see the evolution of the corresponding mode amplitudes. It is striking
that, although the eigenvalues undergo substantial alteration near resonance, the original amplitudes
are largely unaffected as the resonance is crossed.
\label{Snap4}}
\end{figure}

\subsection{Mean Motion Resonances}

A superficially even more dramatic behaviour is shown in Figure~\ref{Snap4}, in which the inner
two planets cross the 2:1 resonance as a result of the tidal evolution of the inner planet. As
the pair approaches the commensurability, the resonance-averaged corrections to the secular modes
become substantial, leading to big changes in  the eigenfrequencies. However,
this behaviour does not result in a substantial change in the long-term qualitative behaviour of the
system, as can be seen from the lower panel of Figure~\ref{Snap4}. This is a consequence of the narrow width of the
resonant region at these low masses, so that the shuffling of the eigenfunctions is a relatively
brief interlude, after which most of the original mode amplitude is returned as the system continues
to evolve past the resonance.

\begin{figure}
\includegraphics[width=84mm]{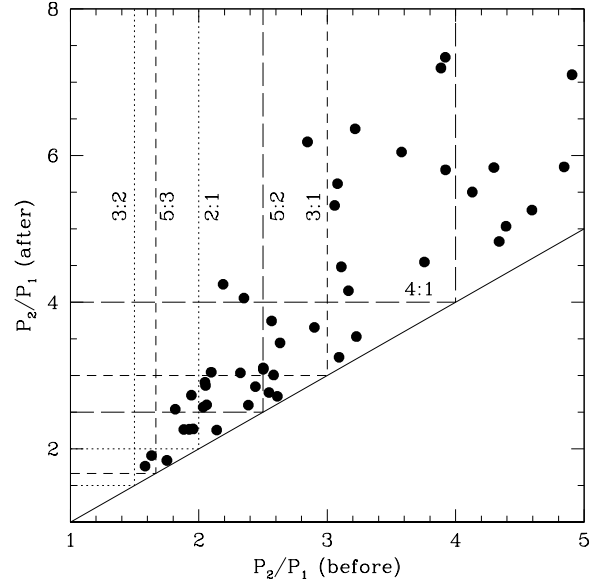}
\caption[Prat.ps]{The filled circles indicate the relationship between the original
period ratio of the inner pair in each system to the final period ratio, after 10~Gyr
of evolution under the influence of tides and secular interactions. Clearly points cannot
be found under the solid line. The lines indicate commensurabilities corresponding to
prominent first order (dotted), second order (short dashed) and third order (long dashed)
resonances. If a point is located to the upper left relative to a given border, then
the planet pair must have crossed that commensurability during its evolution. In some
cases planets have crossed multiple commensurabilities.
\label{Prat}}
\end{figure}

\begin{figure}
\includegraphics[width=84mm]{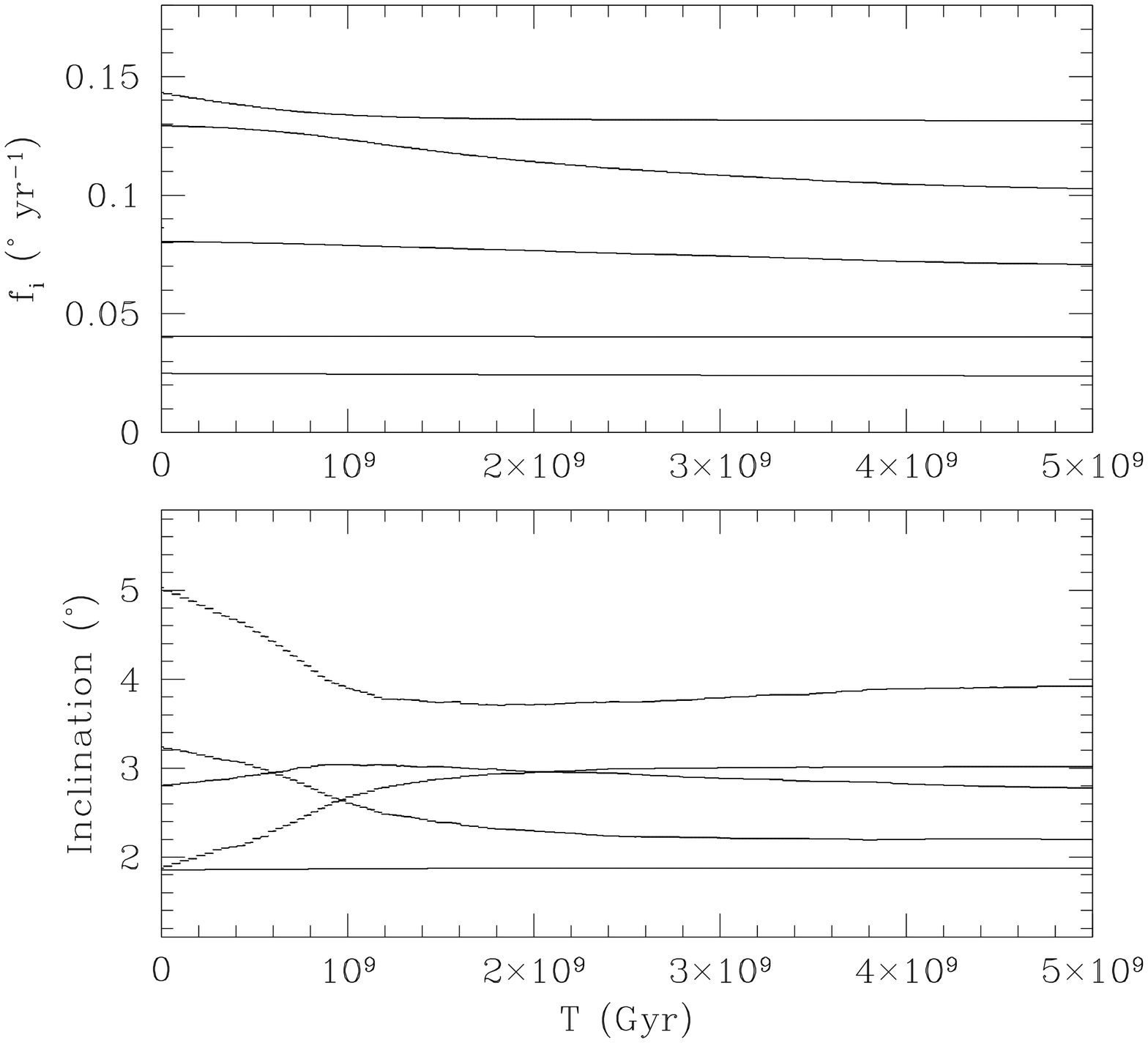}
\caption[Snap5.ps]{The upper panel shows the evolution of the inclination mode eigenvalues
for five modes of a nine planet system. At $\sim 0.8$Gyr, a secular resonance between two
of the modes results in an exchange of amplitude between several modes. The lower panel shows
the evolution of the amplitudes for the five inclination modes with the highest eigenfrequencies.
Although there is interaction, the transfer of amplitude is not sufficient to substantially
alter the qualitative features of the system behaviour.
\label{Snap5}}
\end{figure}

The innermost planets in each system move the most, so the period ratio between the first
and second planets in each system is going to increase. Figure~\ref{Prat} shows the change
in the period ratio for the inner pairs in the 50 systems we investigated. We see that
there are several cases in which the period ratio will cross a low order commensurability at some point during the
tidal evolution. However the planets are diverging, and so will not be captured
in the resonance, and our results suggest that the secular effects of the resonance passage do
not substantially alter the quantitative behaviour of the evolution. In particular, there does
not appear to be sufficient evolution to markedly change the transit observability of planets
in these systems.

\subsection{Inclination Evolution}

In the limit of the classical secular perturbation analysis, the gravitational interactions between
the planets also generate a set of normal modes in inclination, which are not directly coupled
to the modes of eccentricity oscillation. That remains the case in our model as the secular
contribution near first order resonances is linear in eccentricity and all inclination contributions
to the disturbing function are second order, and thus assumed to be negligible. 

Furthermore, tidal damping in the planets occurs in the orbital plane and thus does not cause
any evolution in the planetary inclination. Tidal dissipation in the star can potentially
affect the orbital plane, but we assume this to be negligible due to the fact that the planetary
masses considered here are too small to raise a significant tide on the star.

However, there is a small amount of evolution in the inclination oscillations due to the
fact that the semi-major axis of the innermost planet evolves, effectively providing a 
weak coupling between the eccentricity and inclination evolution. In general, the evolution
in the eigenmodes and eigenvalues is small, but
in about 10\%
of cases this evolution is sufficient to generate secular resonances in the inclination modes
and a certain amount of transfer of amplitude from one mode to another. An example is shown
in Figure~\ref{Snap5}, but we see that the level of adjustment is not sufficient to lead to
qualitative changes in the system appearance.

\begin{figure}
\clearpage
\includegraphics[width=84mm]{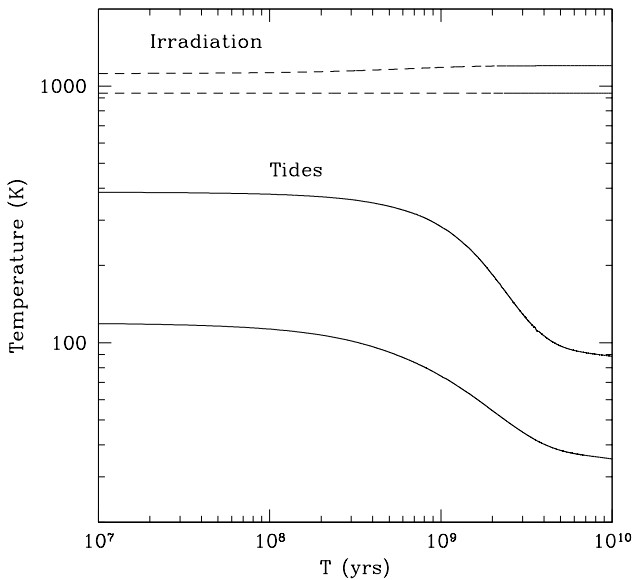}
\caption[Heat.ps]{The solid curves indicate the temperature maintained on the innermost two planets of
the system shown in Figure~\ref{Aevol}
if maintained  by the tidal dissipation. The dashed lines show the equivalent temperature maintained
by the absorption and reradiation of light from the central star. We see that tidal effects are clearly
not the dominant heat source for these planets.
\label{Heat}}
\end{figure}

\subsection{Heating}

The fact that secular effects can extend the
 tidal evolution of an inner planet means that more energy is dissipated. Could this have
an effect on the planet structure? Barnes et al. (2010), for instance, suggest that CoRoT-7b could be
substantially affected by tidal dissipation.
 Figure~\ref{Heat} shows the energy dissipated as a function of time
in the innermost planets of our illustrative example in Figure~\ref{Aevol}. We see that the effective
'tidal equilibrium temperature' is always a factor of several below the equivalent equilibrium maintained
by irradiation from the central star. As such, we do not expect it to have a significant effect.
Such behaviour is ubiquitous in our sample.
The difference between the cases treated here and CoRoT-7b is that the planets in our systems are factors of several
less massive and usually more distant as well.

\subsection{Secular Chaos}

The strong gravitational interactions that result from resonances, either
from mean motion or secular eigenvalue commensurabilities, can potentially
influence the architecture and stability of a system. 
 Although the secular interactions are linear in the approximation
used here, higher order non-linear effects can potentially result in
the transfer of amplitudes from one mode to another, a phenonemon termed
`secular chaos' by  Wu \& Lithwick (2011) and one that may contribute to
the long-term instability of planetary systems, including our 
own (e.g. Laskar 1996).

Although a detailed examination of such questions is beyond the scope of
this calculation, we are in a position to provide an initial estimate of
the susceptibility of these model systems to such evolution. Laskar (1997) introduced
the concept of the quasi-conserved angular momentum deficit, which we will term $\Lambda$, where
\begin{equation}
\Lambda = \sum_{k=1,n}^N \lambda_k = \sum_{k=1,n}^N m_k \sqrt{a_k} \left( 1  - \sqrt{ 1 - e_k^2} \cos i_k \right)
\end{equation}
and $m_k$,$a_k$,$e_k$ and $i_k$ are the planetary mass, semi-major axis, eccentricity and inclination.

If the secular modes are allowed to transfer angular momentum between one another by higher order
interactions, then the individual contributions $\Lambda_k$ may vary, and potentially allow the eccentricity
of a given planet to evolve to higher values. As an initial estimate of the susceptibility of a system to
instability, we can, for each neighbouring pair, ask what value of eccentricity is required for the two orbits
to cross, assuming co-planar orbits. This places a certain critical requirement on individual $\Lambda_k$, which
can be realised if the value for the entire system $\Lambda > \Lambda_{k,crit}$. In effect, if there is a sufficient deficit in the the
system as a whole, the transfer of a substantial portion to a single planet can give that planet an eccentricity
that will cause orbits to cross and potentially engender an epoch of planetary scattering and orbital instability.

\begin{figure}
\includegraphics[width=84mm]{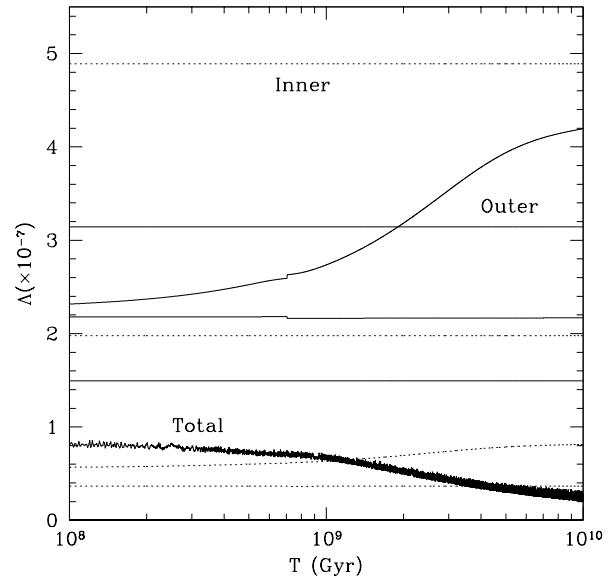}
\caption[AMD.ps]{The solid curves indicate the critical $\Lambda$ required for a specific planet to
have a large enough eccentricity to cross a circular orbit with the semi-major axis of its exterior neighbour.
The dotted curves indicate the value required to cross the orbit of an interior neighbour. The thick curve indicates
the evolution of the $\Lambda$ for the system as a whole, under the action of tidal dissipation. The critical
value associated with pairs involving the inner planet also evolves, as the planet moves inwards. As an example,
the critical values show a small discontinuity at $0.7$~Gyr, associated with a 2:1 resonance crossing. As noted
in the text, 
the tidal evolution of the system reduces the global $\Lambda$ below the critical values on timescales of 1.2 Gyr and 4.5 Gyr respectively.
\label{AMD}}
\end{figure}

However, this is rarely the case in the systems we study here. Examination of both the collective $\Lambda$ and
the required critical values of $\Lambda_k$ for all pairs shows that only $\sim 10\%$ of the systems here have
any pairs that are close enough to be potentially unstable in this manner. Furthermore, the tidal damping reduces
the global $\Lambda$ value, so that most of the potentially unstable systems are rendered stable within 1~Gyr.
Figure~\ref{AMD} illustrates this for the most unstable system in our sample, in which there are initially two
neighbouring pairs that could potentially be destabilised by secular chaos. However, the tidal evolution of the system
reduces the global $\Lambda$ below the critical values on timescales of 1.2~Gyr and 4.5~Gyr respectively. Unless the
nonlinear interactions are strong enough to destabilise the system on this timescale, we do not anticipate orbital
instability. While these arguments are not conclusive they do suggest that the systems studied here are secularly
stable 
unless they are substantially perturbed by an external influence that raises eccentricities above those obtained
from the assembly calculations of Hansen \& Murray (2013).

\subsection{Tidal Strength}
\label{Q}

The strength of the actual tidal dissipation in rocky planets is still somewhat uncertain. Our default
normalisation is based on traditional estimates of dissipation on Earth, but the precise physics that
determines this can potentially be different on extrasolar terrestrial planets. As such, we have recalculated
the evolution above using two alternative tidal strengths, one that is ten times
larger than our nominal calibration ($Q'=1$), and one which is ten times weaker ($Q'=100$).

We find little qualitative change in the overall system behaviour. As expected, less circularisation is found
in cases of high $Q'$ and more circularisation is found in the case of low $Q'$. Furthermore, in systems such
as that shown in Figure~\ref{Snap2}, the nature of the eigenvalue evolution doesn't change -- the decoupling
of the inner planet simply occurs earlier or later, depending on the strength of dissipation. Similarly, resonant
interactions like that in Figure~\ref{Snap3} follow the same pattern except that they occur earlier or later.
The level of eccentricity damping increases as Q decreases, but the
 amount of inward movement is not significantly larger in the Q=1 case than the Q=10 case.

\section{Discussion}

\subsection{The reach of Tidal Circularisation}
\begin{figure}
\includegraphics[width=84mm]{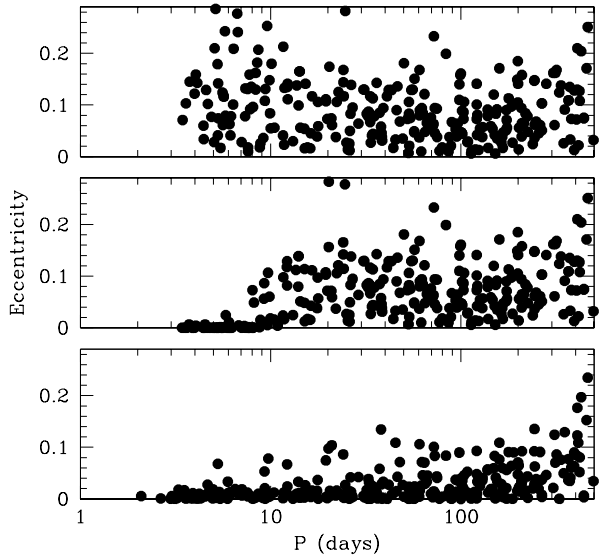}
\caption[Pe.ps]{These panels show the effects of tidal dissipation
and secular coupling in model planetary systems around a $1 M_{\odot}$ star.
The upper panel shows the initial conditions, the
middle panel the effect of single planet tidal evolution, and the lower panel
the consequences of tidal+secular evolution.
\label{Pe}}
\end{figure}

The propagation of tidal damping through a planetary system by the
secular interactions has the effect of dramatically extending the reach
of tidal circularisation. Figure~\ref{Pe} shows the period-eccentricity
relation for our full ensemble of 50 model systems, at three levels of
approximation. The upper panel shows the initial conditions for this
calculation (the end result of the assembly simulations in Hansen \&
Murray 2013). The middle panel shows the consequence of 10~Gyr of tidal
evolution using the equations~(\ref{Tideq}) but applied to each planet
individually without any secular interactions. We see that the 
circularisation period for a single planet system is $\sim 8$~days using
$Q'\sim 10$.
In the lower panel, we show the outcome of 10~Gyr of evolution within
the context of the same model. We see substantial damping
of eccentricities out to orbital periods $\sim 100$~days. 

This extended tidal reach may help to explain some of the inferences
made on the basis of Kepler data. The observed asymmetry of the period
ratios of neighbouring pairs (Lissauer et al. 2012; Fabrycky et al. 2012)
can potentially be explained by the resonant repulsion of near-commensurable
pairs evolving under the influence of tidal dissipation (LW12, BM13).
 However, many of these systems are observed
with orbital periods in the range 10--50 days, where traditional estimates of
tidal damping suggest this is not possible (Lee, Fabrycky \& Lin 2013). 
The estimate presented here suggests that the effects of tidal damping can indeed extend out
to these distances, and that the effects of resonant divergence may
still play a role on these scales. Although the original derivations mostly assumed the traditional
relationship between the damping of eccentricity and semi-major axis to conserve
angular momentum, the qualitative behaviour is unchanged if we allow for eccentricity
damping in the absence of any semi-major axis change.

\begin{figure}
\includegraphics[width=84mm]{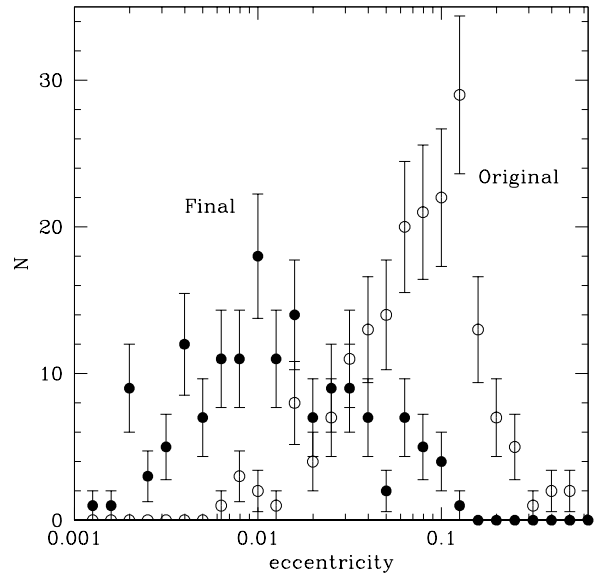}
\caption[Edis.ps]{The open circles show the distribution of eccentricities
at the end of the assembly simulations, before any tidal evolution is applied.
The filled circles show the distribution of eccentricities after 10~Gyr of
evolution under the action of tides and secular perturbations. Only planets with
final orbital periods $<100$~days are shown.
\label{Edis}}
\end{figure}

\begin{figure}
\includegraphics[width=84mm]{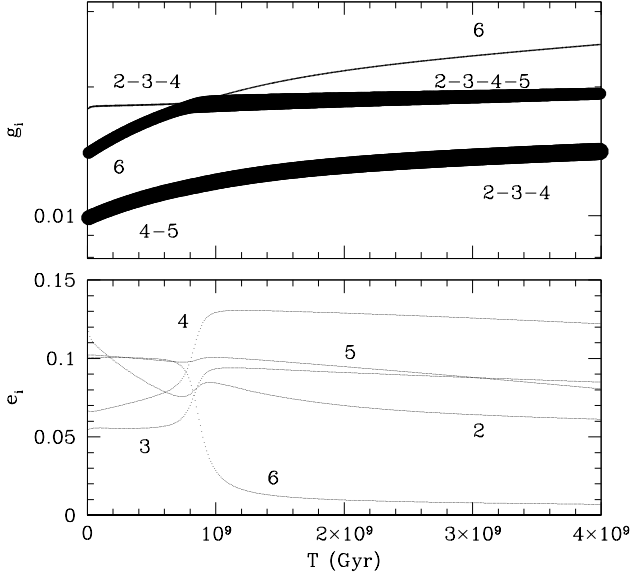}
\caption[Squeeze.ps]{The upper panel shows the evolution of three of the
eigenvalues of a six planet system. The width of the curves is scaled relative
to the mode amplitude and they are labelled by the planets which contribute
at least 10\% to the eigenvector amplitude. Thus, we see that the crossing of two eigenvalues at
$\sim 0.8$Gyr transfers substantial amplitude from an outer mode to an inner mode.
The lower panel shows the evolution of the corresponding individual planet eccentricities.
Here the resonant crossing is manifested by a rapid drop in the eccentricity of the
outermost planet (6) and the corresponding excitation of planets 2,3,4 and 5.
\label{Squeeze}}
\end{figure}

Indeed, eccentricity constraints based on transit timing variations (Lithwick, Xie
\& Wu 2012; Wu \& Lithwick 2013) suggest that many of the observed systems have eccentricities
$<0.01$, which may again be explained on the basis of Figure~\ref{Pe}. They also
observe that a minority appear to exhibit anomalously large eccentricities. Given
that `large', in this context, is more than a few $\times 0.01$, we can see that
such a tail is also potentially explained by the final distribution of our simulations.
Figure~\ref{Edis} shows the final model distribution of eccentricities for all planets
with orbital periods $<100$~days, compared to the original distribution. We see
that the final median eccentricity is indeed $\sim 0.01$, but with a small tail
of values extending up to $\sim 0.1$.
This tail of remnant eccentricities has several potential sources. The simplest is
that a fraction of the
 systems have the inner planet
 particularly poorly coupled to the other planets, so that the damping is weaker.
In the opposite limit, secular resonance can transfer substantial amplitudes from
one mode to another, potentially re-exciting the eccentricities of some of the
close-in planets. An example of this is shown in Figure~\ref{Squeeze}. In this case
the resonant crossing transfers substantial amplitude from a mode dominated by the outermost planet to
several of the interior planets. The newly excited mode is also poorly coupled to the
innermost planet, resulting in a slow subsequent damping and finite eccentricities
for several~Gyr. The inverse is also possible, as seen in Figure~\ref{Snap3}. In that
case, the coupling to a distant planet maintained an inner eccentricity for several
Gyr, until secular resonance drained the amplitude from that mode and circularised
the inner planet. 

\begin{figure}
\includegraphics[width=84mm]{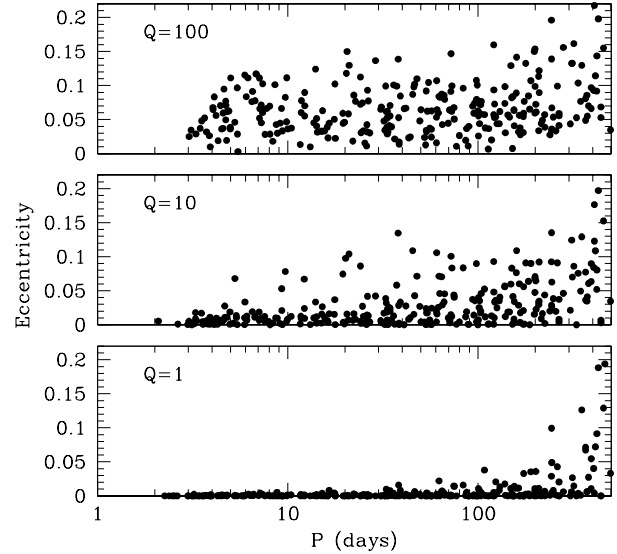}
\caption[Qcomp.ps]{The upper panel shows the final period-eccentricity relation
after 10~Gyr if the strength of tidal dissipation is characterised by Q=100. The
middle panel shows the result for our nominal normalisation Q=10, and the lower
panel shows the result if the dissipation is stronger than the nominal calibration
(Q=1).
\label{Qcomp}}
\end{figure}

The eccentricity distribution is also a function of the strength of tidal dissipation,
as shown in Figure~\ref{Qcomp}. For weaker dissipation, the eccentricities are only
moderately damped, and are probably too large to match the observations. If the
dissipation is much stronger than our nominal calibration then the eccentricites
are even more strongly damped and we see essentially no eccentricities $>0.05$ for
periods $< 100$~days. Taken together, these results suggest that the observations
of Lithwick and collaborators can be explained by tidal dissipation rates of order those in
the Earth, damping an initial eccentricity distribution similar to that seen in
assembly simulations.

\begin{figure}
\includegraphics[width=84mm]{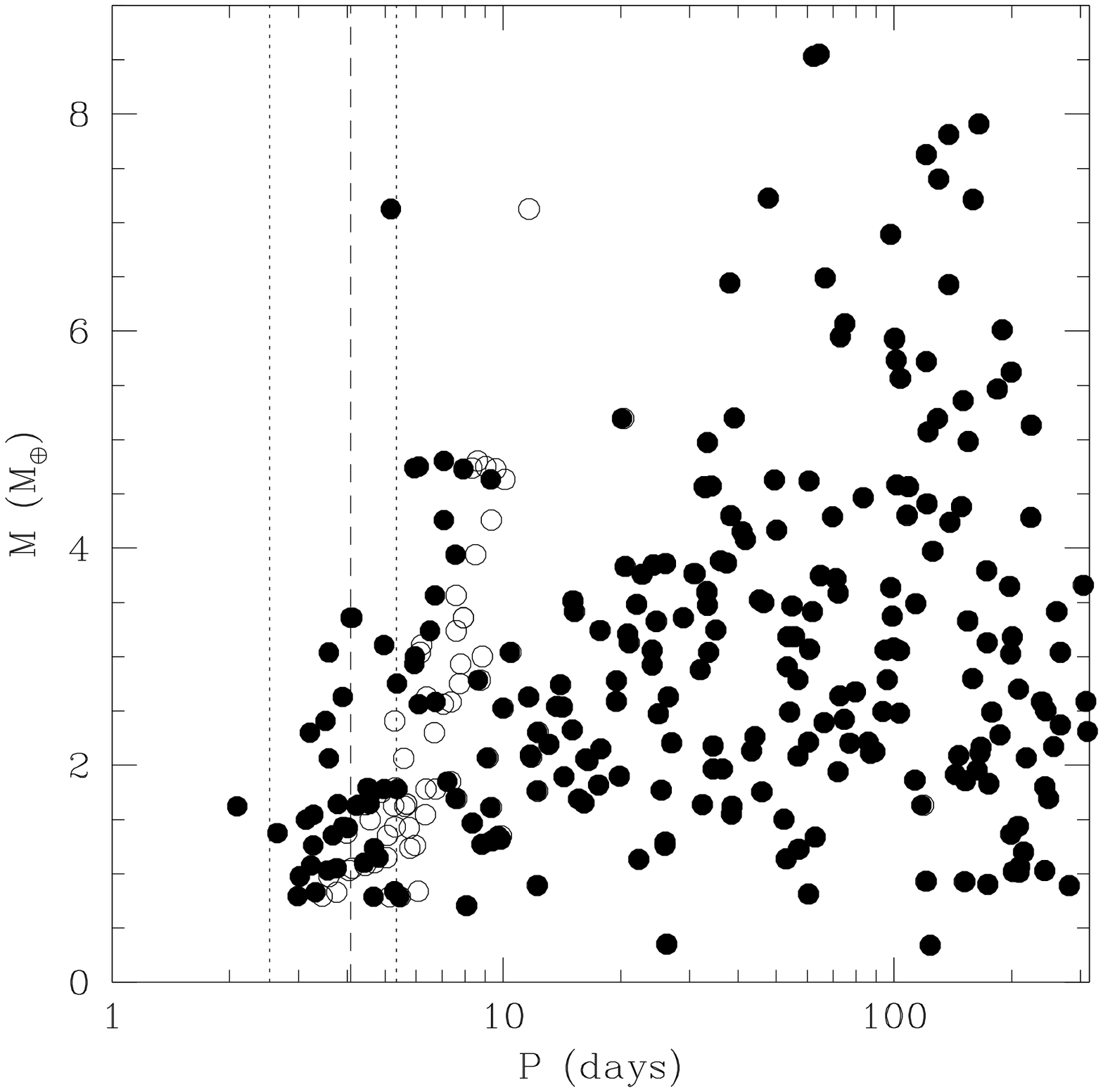}
\caption[Ma.ps]{The open circles show the original masses and orbital periods
for the ensemble of planets in our G~star simulations. The filled circles are the
same systems plotted after 10~Gyr of evolution under the influence of tides and
secular perturbations. The lack of visible open circles at periods $>10$~days
indicates that there has been very little radial motion on these scales. The
vertical dotted lines indicate the estimated sublimation radius for large iron
grains and low density silicate grains respectively (Swift et al. 2013).
The vertical dashed line indicates the inner edge of the initial disk from which
the assembly of the planetary system was calculated in HM13.
\label{Ma}}
\end{figure}

The large scale eccentricity damping also helps to explain the distribution
of observed transit duration ratios. Fabrycky et al. (2011) introduced this
measure to quantify the ratio of impact parameters for neighbouring pairs
of transitting planets. This quantity is a measure of both the dispersion
in inclinations within a planetary system as well as the distribution of
eccentricities. Hansen \& Murray (2013) found that their simulations matched
the observed distribution well if they used the inclination dispersion from
the simulations, but assumed that the orbits were circularised out to semi-major
axes $\sim 0.25$AU or larger. The tidal damping here provides a rationale
for that assumption, because the eccentricities are damped by the tidal dissipation
but the inclinations are not (since planetary dissipation occurs within the orbital
plane, it does not affect the inclination).

\subsection{Migration to short orbital periods}

The radial migration of the inner planets can also potentially explain the presence
of planets on orbits inside the nominal sublimation radius for small bodies. Figure~\ref{Ma}
shows the distribution of period and mass for the ensemble model planets both
before and after tidal evolution. The inner edge of the original distribution is located
at orbital periods $\sim 4$~days, but this shifts in as far as 2~days, depending on
the particular configuration. The migration is not strong enough to drive planets into
the star as has been hypothesized by Mardling \& Lin (2004), because the planetary masses
are not big enough to excite eccentricities to the required levels. However, it is enough
to move planets interior to locations where the nominal building blocks might not survive
during the early stages of the assembly process. Swift et al. (2013) estimate the sublimation
radius for grains and claim that some planets must have migrated inwards at late times because
they could not assemble from dust grains that would sublimate at such locations. The distribution
in Figure~\ref{Ma} indicates that tidal evolution is  capable of driving an inward migration
across such a sublimation barrier after the planet has assembled.

\subsection{Interaction with Giant Planet Systems}

The dense modal structure of these compact planetary systems offers
many possibilities for interesting secular resonant interactions. We
have already observed in \S~\ref{Modes} that we can find secular resonance
interactions even between sub-components of the compact planet systems.
When we consider the interaction with an external giant planet system, there
are even greater possibilities.

The presence of only a single planet can produce substantial effects. As an
experiment, we added a single giant planet at 3~AU to the example system shown
in Table~\ref{Sample}, with a Jupiter-like eccentricity of 0.05. The addition
produces shifts in the eigenvalues, but nothing dramatic unless the planet mass
is $\sim 2 M_J$. For this mass, the presence of the giant planet shifts the 
fifth eigenvalue into position to cross the third eigenvalue at $\sim 2 \times 10^9$~years,
which pumps eccentricity into the interior system.
The consequence of this  is to accelerate the tidal evolution because
the periastron of the inner planet is reduced and the tidal dissipation increased. Repeating this experiment with
other systems produces similar results. Thus, even the presence of a single giant planet of the appropriate mass 
can drive the evolution of the inner planets further than they would have evolved otherwise.

\begin{table*}
\centering
\begin{minipage}{140mm}
\caption{Known Pairs of giant planets on wide orbits
\label{GiantPairs}}
\begin{tabular}{@{}lccc@{}}
\hline
Name & a & Mass & e  \\
    & (AU) &($M_{J}$) & \\
\hline
 HD~37124 & 1.64 & 0.62 & 0.14 \\
          & 3.19 & 0.68 & 0.20 \\
 $\mu$~Ara & 1.511 & 1.67 & 0.27 \\
           & 3.78 & 1.18 & 0.46 \\
HD~183263 & 1.51 & 3.67 & 0.36 \\
          & 4.35 & 3.57 & 0.24 \\
\hline
\end{tabular}
\end{minipage}
\end{table*}

\begin{figure}
\includegraphics[width=84mm]{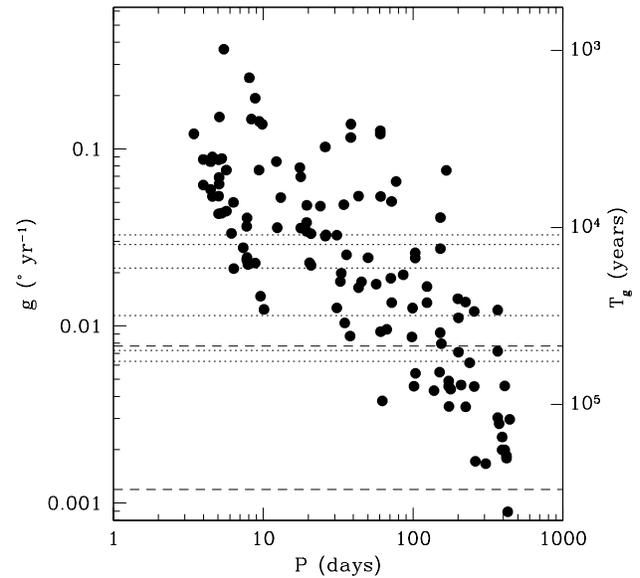}
\caption[gscan2.ps]{The filled circles indicate the eigenfrequencies of modes
in our rocky planet systems around G stars, as in Figure~\ref{gscan}, plotted against
the orbital period of the dominant planet in each mode. The horizontal dotted lines
indicate the eigenfrequencies of the giant-planet-pair systems given in
Table~\ref{GiantPairs}. The dashed lines indicate the two highest eigenfrequencies
associated with the giant planets in our own solar system. The largest of these
is a Jupiter-Saturn mode, while the smaller contains significant contributions
from Jupiter, Saturn and Uranus. The amount of overlap between the compact and giant
planet systems suggests that a rich dynamical coupling is possible.
\label{gscan2}}
\end{figure}

A better comparison with our own Solar system can be achieved by comparing
the eigenfrequencies of the compact systems with those of known systems of
multiple giant planets. Table~\ref{GiantPairs} lists three pairs of giant
planets from the compilation of Wright et al. (2009), chosen so that we have
at least two planets with semi-major axis $> 1.5$AU, which therefore make
plausibly stable exterior systems for our model compact systems. Figure~\ref{gscan2}
shows the eigenfrequencies (dotted lines) of these pairs, compared with the
ensemble of eigenfrequencies shown in Figure~\ref{gscan}. Also shown (dotted lines)
are the two highest eigenfrequencies of the solar system giant planet modes,
quoted in Murray \& Dermott (1999). Although it requires a little fine tuning
to match up the eigenfrequencies in any given system, the density of overlap
between the ensembles suggests that there is much potential for overlap and
interaction if there is even a little bit of drift in the frequencies during
the formation process -- either due to the evaporation of the nascent gas 
disk (Ward 1981) or due to planetesimal-driven evolution amongst the outer
planets (Thommes, Duncan \& Levison 1999; Tsiganis et al. 2005).

\begin{figure}
\includegraphics[width=84mm]{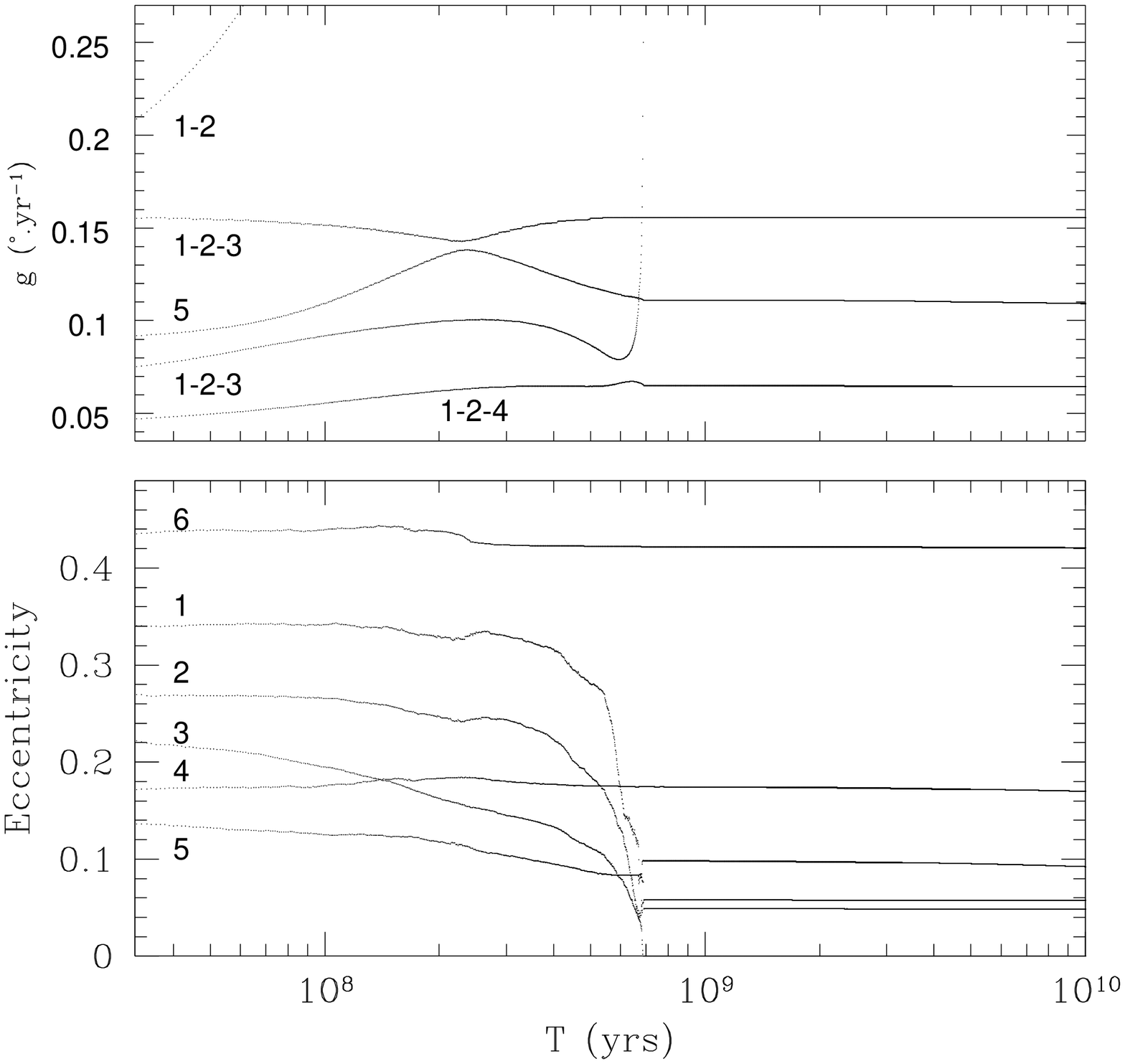}
\caption[Pump2.ps]{The upper panel shows the evolution of the eigenvalues of the
eight planet system composed of the six terrestrial planets in Figure~\ref{Snap1}
plus the two giant planets of HD183263. Three eigenvalues are
not shown as they are sufficiently separated from the others that they don't evolve. The curves
are labelled with the numbers of the planets (counting outwards) which contribute at least 10\% to the
amplitude of the corresponding eigenvector amplitude (at the beginning of the evolution).
The lower panel shows the evolution of the root mean square eccentricity of each of the
6 inner planets, with substantial eccentricity changes associated with several of the
resonance crossings. As before, the net effect of this is often to accelerate the tidal
evolution.
\label{Pump2}}
\end{figure}

Such interactions have been discussed by several authors in the context of our
own solar system. The low eccentricities of the terrestrial planets
suggest that such interactions must have been weak unless there was substantial
dissipation at later times (Ward 1981; Agnor \& Lin 2012) and could also influence
the original assembly of the planets (Levison \& Agnor 2003; Nagasawa, Lin \& Ida, 2003; Nagasawa, 
Lin \& Thommes 2008; Raymond, Barnes \& Mandell 2008; Thommes, Nagasawa \& Lin 2008). 
To illustrate the potential for such interactions with
the model systems discussed here, Figure~\ref{Pump2} shows the evolution of the system
shown in Figure~\ref{Snap1}, when paired with the giant planet pair orbiting HD183263. These giant
planets are both massive, and with substantial orbital eccentricity, so they will provide
greater perturbations than the giant planets of our own solar system. Indeed, we see that
the tidal evolution of this system exhibits a substantial amount of interaction, with 
multiple resonance crossings. These are driven by both the modifications to the eigenvalues
introduced by the additional planets 
and the greater tidal evolution because of the larger secular pumping of the eccentricities.
Indeed, the inner planet ends up at 0.011~AU, instead of at 0.046~AU
when evolved without the giant planets. Of course, the eccentricities in this case reach
values that are probably too large for an accurate treatment with classical linear secular
theory, but this does indicate the prospects for more dramatic evolution when such compact
planetary systems are paired with the giant planet populations already observed. In particular,
the angular momentum deficit $\Lambda$ in such systems is high enough to open up the possibility
of secular chaos and further dynamical evolution. This echoes the situation in our own solar
system, where the secular effects that imperil Mercury are a consequence of the interaction
with the precession of Jupiter (Laskar 1994; Batygin \& Laughlin 2008; Laskar \& Gastineau 2009; Wu \& Lithwick 2011).
Such dynamical evolution may help to explain the puzzling excess of low multiplicity systems in the
Kepler database (Hansen \& Murray 2013).

\section{Conclusions}

We have investigated the long-term evolution of compact planetary systems composed of terrestrial-class planets,
under the action of tidal dissipation and secular gravitational interactions. We have addressed the difficulty
of fully characterising observed systems by studying the evolution of the model systems discussed in Hansen \& Murray (2013).
In that paper, we demonstrated that these simulations broadly reproduced the observed properties of the Kepler   
multiple planet systems, in terms of multiplicities and orbital spacing, and so their secular interactions should
represent a qualitatively representative sample of those found in the observed systems. 

Our results suggest that, for tidal dissipation similar to that found in Earth-like planets, a variety of
dynamical interactions can ensue, including resonances between different secular modes and the crossing
of mean motion resonances. The secular couplings can also extend the reach of tidal circularisation by
an order of magnitude, circularising the orbits out to  periods $\sim 100$~days in most cases.
However, eccentricities $> 0.03$ remain in a minority of cases, resulting either from the decoupling of
the secular interactions into multiple weakly-connected subsystems or from the late-time crossing of a
secular resonance which can redistribute angular momentum amongst the planets and re-excite the eccentricity
of short period planets. 

The secular pumping of eccentricity also helps to drive the inward migration of the inner planets, resulting
in shorter final orbital periods than would result from individual planet migration. In particular, this
can result in planets with final orbital periods that lie inside the so-called 'dust sublimation radius', because
they assembled from smaller bodies farther out and then were driven in by tidal dissipation enhanced by
secular eccentricity pumping. The closest orbital periods found in these simulations were $\sim 2$~days, but planets could
be driven in further if they interact with more massive planets than considered here. The secular interactions
drive migration of the innermost planet primarily, with the result that the period ratio between the closest and
second closest planets diverge substantially, while the period ratios of planets farther out do not. This may help
to explain the observation of Steffen \& Farr (2013), that the period ratios of the innermost pairs in Kepler systems 
are larger, on average, than those of more distant pairs.

We evolve models composed solely of rocky planets with masses $< 10 M_{\oplus}$, within the
context of classical secular perturbation theory. As such, our calculations represent the linear superposition
of individual modes and cannot probe the possibilities of chaotic evolution due to nonlinear couplings.
However,  the level of
the angular momentum deficit in these systems is low enough that significant dynamical instability may not
occur because, in most cases, orbit crossing cannot be achieved even if all of the system's angular momentum
deficit were concentrated in a single mode. 
It is worth noting, however, that this is the result of the low planetary masses considered. If the assembling
planets capture significant gas mass from the surrounding nebula (e.g. Hansen \& Murray 2012), or are accompanied
by a giant planet system at larger radii, the injection of significant additional amplitude could substantially
modify the system evolution. This is an attractive possibility for explaining the excess of low multiplicity systems
seen in the Kepler sample (Lissauer et al. 2011; Dong \& Tremaine 2012, Fang \& Margot 2012; Hansen \& Murray 2013)
and will be examined in detail in a forthcoming paper. 

Finally, we note that these simulations are chosen because they resemble the statistical properties of the observed
planetary systems. As such, even though the systems are the result of in situ assembly calculations, they can 
qualitatively represent the outcomes of any formation scenario that produces a similar range of masses and separations.
However, the degree of evolution does depend on the initial level of eccentricity in the system. Although not large,
the initial eccentricities from the assembly simulations are definitely non-zero. A system born with zero eccentricity
(perhaps due to dissipation from the gaseous nebula) would evolve less. Of course, if the eccentricities are excited
due to dynamical instabilities that result from the dissipation of the nebula, then the level of subsequent tidal
evolution is likely to be similar to that found here.

\newpage

\appendix
\section{Explicit Formulae for Secular Matrix Elements}
\label{Formulae}

The classical solution to the secular problem for a system of N planets, in the limit of low mass ratio (relative to the
central star) and low eccentricity and inclination, is to convert Lagrange's evolution equations into an eigenvalue
problem of the form
\begin{equation}
 A_{ij} x_j = g \, x_j 
\end{equation}
where the variables $x_j$ can be either the vectors formed by the eccentricity and longitude of pericenter 
$h_j = e_j \sin \bar{\omega}_j$ and $k_j=e_j \cos \bar{\omega}_j$, or the equivalent vectors formed from
the inclination and longitude of the ascending node. In our analysis, the
 matrix elements $A_{ij} = A_{ij} + A'_{ij}$ are a combination of the classical secular perturbations $A_{ij}$ and
an additional contribution $A'_{ij}$ that results from the long-term average of any terms in the disturbing function
that result from proximity to a first-order resonance.

The $A_{ij}$ also include a contribution from relativistic precession because of the close proximity of some of
the planets. Diagonal elements are given by 
\begin{equation}
A_{ii}  = n_i \left(  \frac{1}{4} \sum_{j=1,j \neq i}^{N} \frac{m_j}{m_*+m_i} \alpha_{ij} \bar{\alpha}_{ij} b_{3/2}^{(1)}(\alpha_{ij})
+ 3 \frac{G M_*}{c^2 a_j} \right)
\end{equation}
and non-diagonal terms are given by 
\begin{equation}
A_{ij}  =  - \frac{n_i}{4} \frac{m_j}{m_* + m_i} \alpha_{ij} \bar{\alpha}_{ij} b_{3/2}^{(2)}(\alpha_{ij})
\end{equation}
where $n_i$ and $m_{i}$ are the planetary orbital frequency and mass, $m_*$ is the central object mass and
 $\alpha_{ij} = \bar{\alpha}_{ij} = a_i/a_j$  if $a_i<a_j$ (where $a_i$ is the semi-major axis). If $a_i>a_j$,
then $\alpha_{ij}=a_j/a_i$ and $\bar{\alpha}_{ij} = 1$. The functions $b^{(1)}_{3/2}$ and $b^{(2)}_{3/2}$ are
the Laplace coefficients. These are evaluated over all combinations of i and j.

To evaluate the contribution $A'_{ij}$ from near-resonant interactions, we calculate contributions based on
the proximity of each pair i,j to the 3:2 and 2:1 resonances. 
To first order in the masses, proximity to a first order resonance $k:k-1$ is given by
\begin{equation}
 \upsilon  = \frac{3}{2} \frac{ \left[ (k-1)^2 m_j/m_c \alpha_{ij}^{-2} + k^2 m_i/m_c \right]}{ \left[ (1-k) n_i/n_j + k \right]^2}.
\end{equation}

The diagonal contribution to the secular frequencies is then given by
\begin{equation}
A'_{ii} = \upsilon \sum_{j=1, j \neq i}^N n_i \frac{m_j}{m_c} C_{ij}
\end{equation}
where the quantity $C_{ij} = \alpha_{ij} C_1^2(\alpha_{ij})$ if $i<j$ and 
$C_{ij} = C_2^2(\alpha_{ij})$, where $C_1$ and $C_2$ are defined below. 
The off-diagonal terms are given by
\begin{equation}
A'_{ij} = \upsilon n_i \frac{m_j}{m_c} \bar{\alpha_{ij}} C_1  C_2.
\end{equation}

For the case k=2 (i.e. the 2:1 resonance), the quantities $C_1$ and $C_2$ are
\begin{eqnarray}
C_1 & = & -2 b_{1/2}^{(2)} - \frac{1}{4} \alpha b_{3/2}^{(1)} + \frac{1}{2} \alpha^2 b_{3/2}^{(2)} - \frac{1}{4} \alpha b_{3/2}^{(3)} \\
C_2 & = & \frac{3}{2} b_{1/2}^{(1)} - \frac{1}{2 \alpha^2} + \frac{1}{4} \alpha b_{3/2}^{(0)} - \frac{1}{2} \alpha^2 b_{3/2}^{(1)} + \frac{1}{4} \alpha b_{3/2}^{(2)}
\end{eqnarray}
where the $b_i^{(j)}$ are all evaluated at $\alpha$.
For the case j=3 (i.e. the 3:2 resonance), the quantities $C_1$ and $C_2$ are
\begin{eqnarray}
C_1 & = & -3 b_{1/2}^{(3)} - \frac{1}{4} \alpha b_{3/2}^{(2)} + \frac{1}{2} \alpha^2 b_{3/2}^{(3)} - \frac{1}{4} \alpha b_{3/2}^{(4)} \\
C_2 & = & \frac{5}{2} b_{1/2}^{(2)}  + \frac{1}{4} \alpha b_{3/2}^{(1)} - \frac{1}{2} \alpha^2 b_{3/2}^{(2)} + \frac{1}{4} \alpha b_{3/2}^{(3)}.
\end{eqnarray}
Here we have used the formalism of Malhotra et al. (1989) with the addition of the contribution of the indirect term for the external member of
the pair in the case of
the 2:1 resonance.

Our treatment of resonant crossing measures the width of a resonance by taking the largest of two following libration widths. Murray \& Dermott (1999) 
provide an estimate in the case of a test particle being perturbed by an external planet of mass $m_2$ on a circular orbit, which can be expressed as
a difference $\epsilon = 2 - P_2/P_1$, in terms of inner eccentricity $e_1$,
\begin{equation}
\epsilon \sim \left( 36 \frac{m_2}{m_c} e_1 \right)^{1/2} \left[ 1 + \frac{1}{36 e_1^3} \frac{m_2}{m_c} \right]^{1/2} - \frac{1}{2e_1} \frac{m_2}{m_c}.
\end{equation}
An equivalent estimate can be made for the effect of an internal planet of mass $m_1$ on an external test particle, which can be expressed as
\begin{equation}
\epsilon \sim \left( 27 \frac{m_1}{m_c} e_2 \right)^{1/2} \left[ 1 + \frac{1}{48 e_2^3} \frac{m_2}{m_c} \right]^{1/2} - \frac{3}{8e_2} \frac{m_2}{m_c}.
\end{equation}
The equivalent expressions for the 3:2 resonance, in terms of $\epsilon = 3/2 - P_2/P_1$, are
\begin{equation}
\epsilon \sim \left( 41.7 \frac{m_2}{m_c} e_1 \right)^{1/2} \left[ 1 + \frac{1}{70 e_1^3} \frac{m_2}{m_c} \right]^{1/2} - \frac{1}{2.6 e_1} \frac{m_2}{m_c}
\end{equation}
and
\begin{equation}
\epsilon \sim \left( 29.3 \frac{m_1}{m_c} e_2 \right)^{1/2} \left[ 1 + \frac{1}{100 e_2^3} \frac{m_2}{m_c} \right]^{1/2} - \frac{3}{11 e_2} \frac{m_2}{m_c}.
\end{equation}

Our calculations also include the evolution of the inclination oscillations. In this case, the modifications due to relativity, first order resonances, and
tidal damping do not apply as all of those physical effects are exerted in the plane of the planetary orbit. Thus, the equivalent $B_{ij}$ for the inclinations
are simply the usual classical expression, given here for completeness:

Diagonal elements are given by
\begin{equation}
B_{ii}  = - n_i   \frac{1}{4} \sum_{j=1,j \neq i}^{N} \frac{m_j}{m_*+m_i} \alpha_{ij} \bar{\alpha}_{ij} b_{3/2}^{(1)}(\alpha_{ij})
\end{equation}
and non-diagonal terms are given by
\begin{equation}
B_{ij}  =   \frac{n_i}{4} \frac{m_j}{m_* + m_i} \alpha_{ij} \bar{\alpha}_{ij} b_{3/2}^{(1)}(\alpha_{ij}).
\end{equation}

\end{document}